# Swelling-Induced Stress-Assisted Transfer of Nanodiamond Arrays with a PVA Carrier Tape for Conformal Bio-Integrated Sensing and Labelling

Luyao Zhang[1], Lingzhi Wang[1], Xinhao Hu[1], Yip Tai Nam[1], Mingzhe Sun[2], Jixiang Jing[1], Lizhi Xu[2], Yuan Lin[2*], Yong Hou[1*], Zhiqin Chu[1,3*]

1.      Department of Electrical and Electronic Engineering, The University of Hong Kong, Pok Fu Lam, Hong Kong.
2.      Department of Mechanical Engineering, The University of Hong Kong, Pok Fu Lam, Hong Kong.
3.      School of Biomedical Sciences, The University of Hong Kong, Hong Kong.

* Corresponding authors:
Prof. Dr. Yuan Lin, Email: ylin@hku.hk
Dr. Yong Hou, Email: houyong@eee.hku.hk
Prof. Dr.  Zhiqin Chu, Email: zqchu@eee.hku.hk

## Abstract

The conformal integration of nitrogen-vacancy (NV) center nanodiamond arrays onto soft, hydrated, curvilinear biological interfaces remain a fundamental challenge for in vivo quantum sensing and imaging. Conventional transfer techniques often fail due to reliance on high temperature, corrosive chemicals, or mechanical peeling, leading to pattern damage, low fidelity, or poor biocompatibility. Here, we report a transfer strategy utilizing polyvinyl alcohol (PVA) carrier soluble tape, enabling rapid, residue-free, high-fidelity transfer of nanodiamond patterns onto diverse biointerfaces. The success of this method is rooted in a unique "hydrate-soften-expand-self-peel" mechanism of the soluble tape with PVA backing. In situ mechanical tracking reveals non-uniform PVA swelling upon hydration generates transient local normal and shear stresses at the interface. These stresses delaminate the tape within 3 minutes at room temperature while promoting adhesion of the nanodiamond array to the substrate. In contrast, conventional water-soluble tapes with composite structures undergo passive dissolution and collapse, causing residue contamination and reduced efficiency. Leveraging this mechanism, we achieve conformal patterning on ultra-soft hydrogels (~0.6 kPa) and highly curved bio-surfaces (hair, 100 μm$^{-1}$). Additionally, we demonstrate a dual-identity verification system integrating data storage and physical unclonable functions on a hydrogel contact lens. This work provides a versatile tool for bio-interface engineering and a general framework for gentle, efficient transfer of functional nanomaterials.

## 1. Introduction

In recent years, nitrogen-vacancy (NV) centers in nanodiamonds (NDs) have emerged as a versatile platform for advanced biological sensing, [1-5] enabling high-sensitivity detection of magnetic fields, [6-8] temperature, [9-12] and other physicochemical parameters at the micro- and nanoscale. [13-15] While single-ND probes have shown promise, patterned ND arrays are attracting widespread research interest due to their capability for spatially resolved, large-area, and real-time biosensing. For example, ND arrays embedded in elastic PDMS substrates enable the mapping of cellular traction forces across an entire cell adhesion area, providing insights into mechanical signaling within complex cell-matrix interactions. [16] As well as this, ND-embedded boronic acid hydrogels deployed as skin patches facilitate continuous, real-time glucose monitoring for diabetes management. [17] To realize these promising applications, ND arrays must conformally and stably interface with biological substrates such as tissues, skin, and biomaterials. However, the inherent softness, hydration, and curvilinear nature of these interfaces often cause conventional transfer methods to fail, resulting in pattern fracture, delamination, or substrate damage. This fundamental incompatibility severely limits the use of patterned NDs in *in situ* and real-time biosensing.

Current techniques for fabricating nano/micro-material structures, including transfer printing methods that move patterns between substrates using either solid or liquid carriers, are well-established for creating high-density patterns with precise particle positioning. Solid carriers are typically based on elastomers [18] or adhesive tapes, [19] while liquid carriers employ thinner, more flexible supports [20] or the liquid itself as the medium. [21] However, their reliance on flat substrates and processing conditions involving elevated temperatures, corrosive chemicals, or high-energy exposure renders them unsuitable for delicate biological interfaces. To overcome the limitation of non-planar patterning, degradable and morphologically adaptive transfer strategies have been recently developed. A notable example is the reflow transfer technique, [22] which uses a sugar-based carrier to conformally transfer micro-nano structures onto 3D curvilinear surfaces. However, this method requires multi-hour thermal curing at temperatures above 80°C, which can induce osmotic stress and structural dehydration in hydrated biological substrates, thus limiting its use for sensitive live-cell or in-tissue applications.

To bridge the gap, we have developed a water-soluble tape transfer printing (WTT) method, which leverages the tape's exceptional conformability to complex geometries and complete water solubility/detachability to enable rapid (< 3min), residue-free and high-fidelity transfer of nanodiamond arrays on various biointerfaces. Crucially, we discovered that the transfer efficacy is fundamentally governed by the tape's internal architecture and its dynamic response to water. Unlike conventional dissolution-based release, our approach exploits the functional bilayer architecture of polyvinyl alcohol (PVA)-based tapes, which undergo a rapid hydration-swelling process. This swelling generates transient interfacial compressive and shear stresses that actively press the ND array against the target substrate while promoting the tape's self-delamination. This "force-assisted" release mechanism ensures both high transfer efficiency and a pristine, residue-free interface. We systematically investigated substrate interfacial properties to optimize transfer efficiency across diverse materials, including hydrogels, microstructured polydimethylsiloxane (PDMS), hair, and biomimetic substrates. The method supports dual identity verification for wearable devices, incorporating information storage and anti-counterfeiting functionalities. Furthermore, WTT enables wafer-scale patterning, paving the way for high-performance, bio-integrated sensing platforms.

## 2. Results

### 2.1 Swelling-induced stress-assisted release and transfer

Achieving high-fidelity transfer of nanomaterial patterns onto soft, hydrated, and topologically complex biological interfaces requires a release mechanism that is both gentle and effective. To address this, we propose a WTT strategy that utilizes the unique properties of water-soluble tapes: excellent conformability to non-planar surfaces and the ability to be completely removed in wet environments. As illustrated in **Figure 1**a, the WTT process involves four key steps: (1) Fabricating ND patterns on a silicon wafer; (2) Lifting the patterns using a water-soluble tape; (3) Conformally attaching the ND array-laden tape onto the target substrate; (4) Immersing the assembly in water to dissolve the tape, resulting in the pristine transfer of the ND array.

Water-soluble tapes commercially available fall into two structural categories. The first is bilayer-swelling tapes (e.g., 3M 5414 Water-Soluble Wave Solder Tape) consist of a pure PVA backing layer (~ 33 µm thick) laminated with a thin PVA-based adhesive coating (~ 20 µm thick). This design creates a functionally segregated system where the backing serves as a swelling-active structural scaffold while the adhesive layer mediates ND attachment. The second is composite-dissolution tapes embed non-dissolvable fibers (e.g., cellulose microfibers, 10-50 µm length) directly within the PVA adhesive matrix. While both types are water-soluble, their distinct dissolution pathways produce profoundly different interfacial dynamics and transfer outcomes (Figure 1b, Video 1, Video2).

To elucidate this mechanism, we performed real-time optical microscopy and traction force microscopy (TFM) during tape release on PDMS substrates. Upon water immersion, the Type I bilayer tape exhibited immediate anisotropic swelling (Figure 1e). The PVA layer quickly hydrates and expands (20 s), and then self-peels as a single piece within roughly 60 s. This smooth hydrate–expand–self-peel sequence avoids any abrupt forces that could disturb the pattern. TFM analysis captured the underlying force evolution: at ~20 s, fluorescent ND tracer particles embedded in the substrate displayed radial outward displacement, revealing an ellipsoidal compressive stress field (~ 500-600 Pa) generated by swelling-induced expansion; By 30 s, as the backing reached maximum expansion, this compressive force field dissipated and reversed, showing inward particle recoil as the gel detached (Figure 1c, Figure S2a). In contrast, the composite tape showed only passive dissolution without measurable swelling. Water infiltration slowly fragmented the matrix, dispersing cellulose fibers into the surrounding medium over several minutes (Figure 1f). TFM revealed no significant particle displacement (Figure 1d, Figure S2b), confirming the absence of active mechanical forces. These data directly evidence that swelling generates transient normal forces that press NDs onto the substrate while shear components drive autonomous tape delamination.

We further characterized the material transformation of the isolated PVA backing layer during hydration. The swelling ratio rises sharply to ~500% in the first 30 s before plateauing, while the Young's modulus falls from around 60 kPa to nearly 0.5 kPa over the same period (Figure 1g). The backing essentially transitions from a stiff film to a soft, highly deformable gel, which both accommodates substrate topography and produces the compressive stress needed for efficient particle transfer. The mechanistic divergence has a direct and profound impact on transfer performance. Transfer efficiency with the swelling-based PVA tape consistently exceeds 98%, compared with only 50–60% for dissolution-based tapes (Figure 1h). Critically, the swelling mechanism produced clean, uniform ND arrays with no visible defects or residue (Figure 1i, Figure S4a), whereas dissolution tape results in obvious particle loss, gaps, and residual contamination (Figure 1j, Figure S4b, Figure S4c).

Overall, our comparative analysis establishes that a bilayer architecture enables an active, swelling-induced stress-assisted transfer mechanism. This mechanism is superior to the passive dissolution of composite tapes, as it simultaneously enhances particle adhesion through compressive stress and ensures clean, residue-free removal via self-peeling shear forces. Consequently, we adopted this PVA-based bilayer-swelling tape design for all subsequent investigations and applications.

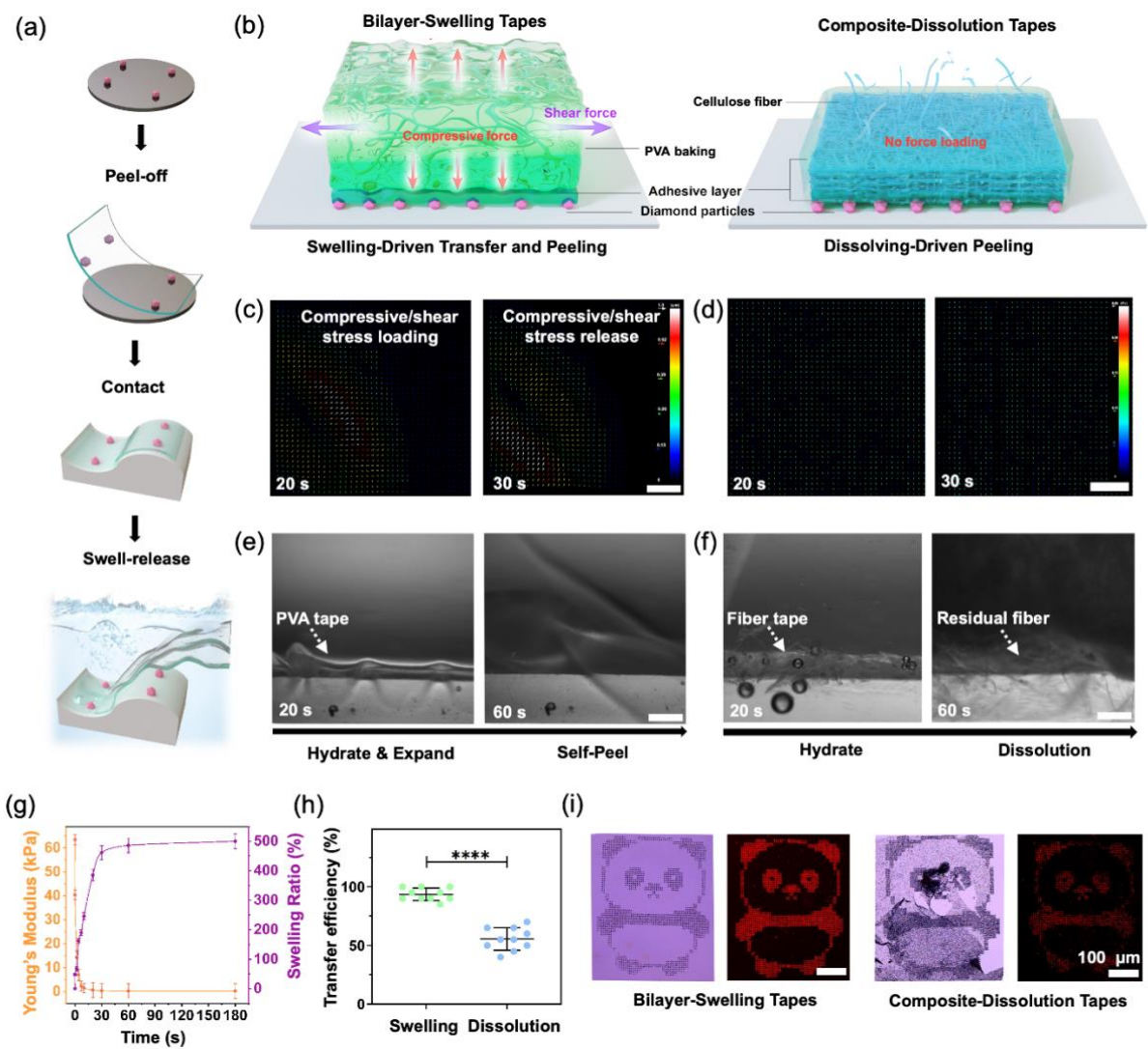

**Figure 1. Swelling-induced stress-assisted transfer mechanism of nanodiamond arrays.** (a) Schematic illustration of the four-step water-soluble tape transfer (WTT) process. (b) Type I (bilayer, swelling-assisted) and Type II (composite, dissolution-based). (c) Traction force microscopy (TFM) analysis showing the transient compressive (outward, ~20 s) and subsequent recoil (inward, ~30 s) displacement fields generated during the swelling and delamination of the Type I tape. (d) TFM analysis showing no force loading during the hydrogel and dissolution of the Type II tape (e) Optical micrographs of the swelling process. (f) Optical micrographs of the dissolution process. (g) Evolution of swelling ratio and Young's modulus during the swelling process of the isolated PVA backing. (h) Statistical comparison of transfer efficiency for the two release mechanisms. (i-j) Optical and confocal photos of transfer outcomes achieved by the swelling and dissolution release mechanisms. Data represent mean ± SD. *P < 0.05, **P < 0.01, ***P < 0.001, ****P < 0.0001 one-way ANOVA. ns, not significant.

2.2 Surface physical and chemical factors influence transfer efficiency of NDs pattern

Biological tissues, such as skin and organs, exhibit high topological heterogeneity, which disrupts uniform contact between the transfer carrier and tissue surface, leading to incomplete pattern transfer. Their stiffness varies significantly, ranging from soft brain tissue (1-3 kPa) [23] to stiffer cartilage (5.7-6.2MPa), [24] resulting in mechanical compliance mismatches that cause uneven pressure distribution and reduced transfer efficiency. Additionally, complex surface chemistries, involving functional groups like amines, carboxyls, and hydroxyls, and biomolecules such as proteins and glycans, [25] generate variable interfacial adhesion through hydrogen bonding, van der Waals forces, and electrostatic interactions, [26] often leading to low transfer efficiency and failure. To address these challenges, we evaluated physical properties, including substrate stiffness, topography and surface chemistry, which influence contact uniformity, pressure distribution, adhesion stability, and ND pattern transfer efficiency on flexible bio-substrates.

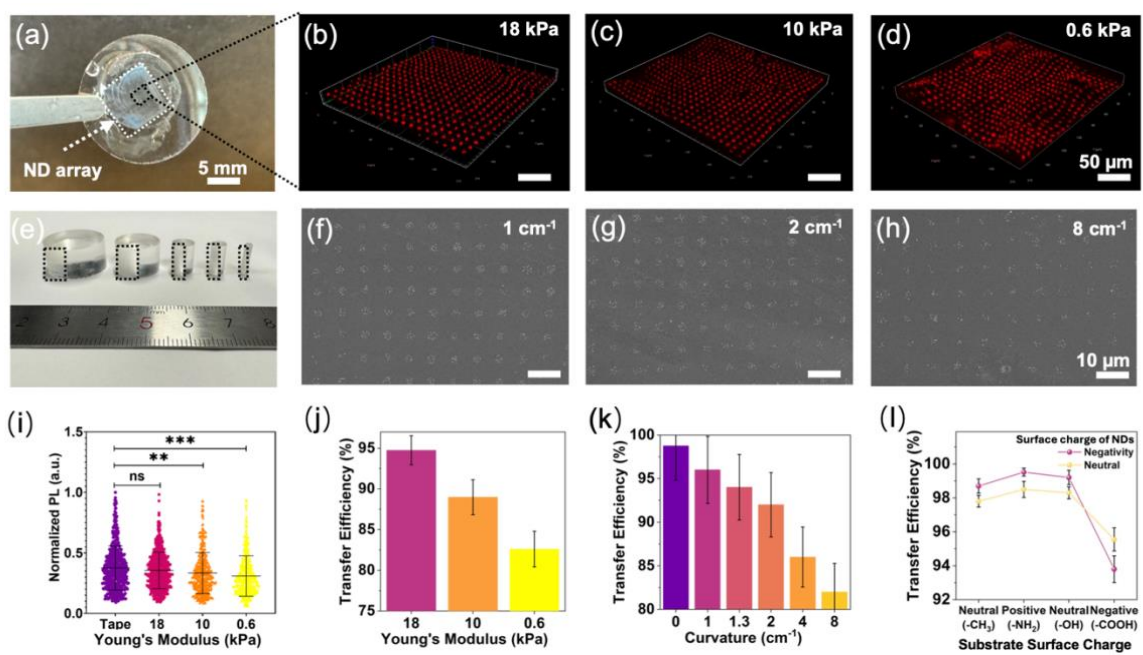

**Figure 2. Nanodiamond transfer efficiency across physical and chemical conditions.** (a) Photograph of nanodiamond array on hydrogel. (b–d) Confocal fluorescence images on hydrogels with varying Young's moduli. (e) Photograph of arrays on PDMS cylinders with different curvatures. (f–h) SEM images of PDMS cylinders. (i) PL intensity statistics. (j) Transfer efficiency calculated from (i) PL intensity statistics on PDMS. (k) Analysis of -H and -COOH nanodiamonds on substrates with $-CH_3$, $-NH_2$, $-OH$, and $-COOH$ groups. Data represent mean ± SD. *$P < 0.05$, **$P < 0.01$, ***$P < 0.001$, ****$P < 0.0001$ one-way ANOVA. ns, not significant.

We first investigated the effect of substrate stiffness on transfer efficiency. Gelatin hydrogel models were selected due to their tunable mechanical properties and biocompatibility, making them suitable for biomaterials and biosensing applications. [27] A stiffness range of 0.6–18 kPa was chosen, covering soft tissues like brain, [23] muscle, [28] and liver, [29] relevant for tissue engineering. To ensure consistent ND pattern transfer, a standardized transfer template was employed, consisting of 4 μm-diameter circles (each containing 15–25 NDs, ~150 nm diameter) in a 10 μm periodic array (Figure S10a). **Figure 2**a shows an ND pattern transferred onto an 18 kPa hydrogel, with confocal images of patterns on 18, 10, and 0.6 kPa hydrogels in Figure 2b–d. Transfer efficiency was quantified using a confocal microscope (NV center fluorescence: excitation wavelength 532 nm, emission wavelength 650 nm), imaging approximately 400–600 circles on the 18, 10, and 0.6 kPa hydrogels. Figure 2i presents a scatter plot of normalized intensities with statistical significance annotated. Transfer efficiency decreased from 94.75 ± 1.80% at 18 kPa to 82.60 ± 2.19% at 0.6 kPa, with a notable decline below 10 kPa (Figure 2j). This inverse correlation between efficiency and compliance is attributed to the increased deformability of softer hydrogels. Upon contact, softer substrates exhibit greater localized deformation, effectively increasing the surface curvature at the micro-scale and reducing the effective contact area. According to contact mechanics principles, [30,31] lower stiffness decreases the contact area and interfacial adhesion energy, thereby hindering pattern transfer. This finding informs substrate optimization for tissue engineering and biosensing applications.

We next investigate curvature effects on transfer efficiency. PDMS cylinders with curvatures ranging from 1 to 8 cm⁻¹ (radii of 0.125–1 cm) were fabricated as mechanically stable models. Figure 2e shows an ND pattern transferred onto a PDMS cylinder (left, dashed box), with SEM images at curvatures of 1, 2, and 8 cm⁻¹ in Figures 2f–h. ND counts in 100 circular regions per curvature were compared to pre-transfer counts on the silicon wafer, using a flat PDMS substrate (0 cm⁻¹) as a control, to calculate transfer efficiency (Figure 2j). The transfer efficiency remained high (>95%) across all curvatures but exhibited non-monotonic variations, with values at 1 cm⁻¹ and 8 cm⁻¹ being comparable yet distinct (Figure 2k). While the flat control achieved the highest efficiency (98.75 ± 3.95%). Transfer efficiency shows a non-monotonic dependence on curvature, [30,31] increasing curvature (decreasing radius) reduces the contact area between the stamp and substrate, decreasing the number of NDs in direct contact with the substrate, which weaken van der Waals interactions, while local stress variations and uneven ND distribution on curved surfaces lead to non-linear fluctuations in adhesion strength. This explains the observed efficiency variation and informs the design of curved substrates for biomedical applications.

Finally, we explored the role of interfacial chemistry. PDMS was selected as the substrate due to its ease of fabrication and excellent potential for surface functionalization. Using functionalization strategies, we modified PDMS surfaces to create biomimetic substrates with varying functional groups and charges, enabling a systematic study of interactions between carboxyl- and hydroxyl-functionalized NDs, commonly used in biosensing (size frequency distribution and zeta potentials detailed in Figure S5 and Table S2; ND preparation methods detailed in the Experimental Section), and these interfaces. Three models were prepared: positively charged -NH$_2$ groups, neutral -OH groups, and negatively charged -COOH groups, with untreated PDMS as a control. Successful modification was confirmed by XPS (Figure S6) and contact angle measurements (Figure S7) showed enhanced hydrophilicity post-modification (modification details in the **Experimental Section**). As shown in Figure 2l and Figure S8, the -NH$_2$ surface demonstrated the highest efficiency due to electrostatic attraction, followed by the -OH surface because of hydrogen bonding, and the -COOH surface illustrated a lowest figure, resulting in surface charge repulsion. These findings reveal that electrostatic attraction strengthens the adhesion between NDs and substrates, thereby enhancing transfer efficiency, while repulsive forces diminish this efficiency. This principle underscores the importance of tailoring surface chemistry to achieve optimal patterning outcomes. The efficacy of surface charge engineering in improving transfer yield highlights their utility in the design of diamond pattern-based biointerfaces.

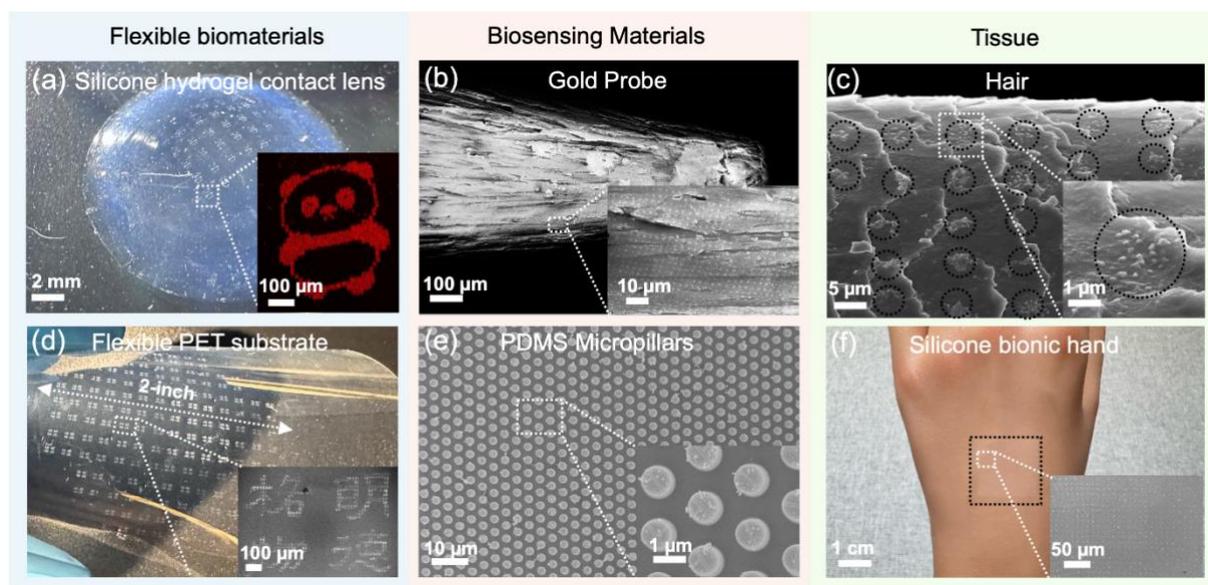

**Figure 3. Versatile transfer of ND patterns onto diverse substrates using the WTT method.**

(a) Photograph (left) and confocal fluorescence image (right) of an ND panda pattern (4 µm diameter circles, 7 µm pitch) transferred onto a silicone hydrogel contact lens. (b) SEM image of an ND array (4 µm diameter circles, 10 µm period) transferred onto a gold probe. (c) Optical micrograph of an ND array (4 µm diameter circles, 10 µm period) transferred onto human hair. (d) Photograph (left) and SEM image (right) of a 2-inch "Ming De Ge Wu" text ND pattern (4 µm diameter circles, 10 µm pitch) transferred onto PET film. (e) SEM image of randomly distributed NDs transferred onto the top surfaces of PDMS micropillars. (f) Photograph (left) and SEM image (right) of an ND array (4 µm diameter circles, 10 µm period) transferred onto a silicone prosthetic hand.

## 2.3 Versatile ND pattern transfer on diverse biomedical substrates using the WTT method

To demonstrate the versatility of the WTT method across key biomedical scenarios, we transfer ND patterns on three representative biointerfaces including flexible biomaterials, biosensing materials, and tissue-related substrates. These categories were selected to encompass the primary challenges in ND-based applications, including adaptability to deformable and hydrated surfaces, precision on intricate microstructures for sensing, and compatibility with irregular, biologically relevant topologies, thereby validating the method's potential to enable scalable, real-time biosensing in wearable devices, neural interfaces, and tissue engineering.

For flexible biomaterials, transferring nanostructures poses significant challenges due to non-planar topology, wet interfaces, and ultra-soft mechanics. We chose a silicone hydrogel contact lens as the model, it provides curved surface with high water content of approximately 50% and Young's modulus of about 0.5 MPa, [32] mimicking a wet, deformable, and non-planar biointerface. We designed a panda pattern formed by nanodiamond arrays (Figure S10b) and transferred it onto the lens. Confocal microscopy in **Figure 3a,** with a magnified view in Figure S9a, demonstrates that the nanodiamond arrays conform seamlessly to the lens's curvature, maintaining uniform particle spacing and alignment without cracks or aggregation, even on the highly deformable hydrogel substrate. This precise transfer highlights the method's ability to preserve intricate details on soft, curved surfaces. To further validate large-area scalability, we extend this approach to larger scale flexible films. We created a complex "Ming De Ge Wu" text array (Figure S10c) and transferred it onto polyethylene terephthalate in Figure 3d, yielding uniform high-fidelity patterns over a 2-inch area.

For biosensing materials, where high-resolution sensing demands the seamless integration of sensor arrays with microscale, three-dimensional device architectures, [33-35] we therefore evaluated the method's performance on specialized probes and arrays. Brain-machine interface probes, for instance, are tiny and sharply curved. To demonstrate our method capability in this context, we transferred standard nanodiamond arrays with 4 µm diameter and 10 µm spacing onto a gold probe tip with 500 μm base diameter and 50 μm tip curvature. SEM images in Figure 3b and Figure S9b show clean, high-fidelity patterns across the curved surface, supporting local magnetic field sensing to track neural activity. Building on this precision for curved microstructures, we also achieved selective transfer of nanodiamond particles onto the tops of PDMS micropillars with 2 μm diameter, 6 μm height and 4 μm interval (Figure 3e). The water-soluble tape touches only the pillar tips, placing particles exactly on the pillar top. SEM confirms nanodiamonds sit solely on the top faces, enabling high-resolution mapping of cellular forces through strain-sensitive fluorescence.

For tissue-related substrates, transfer failures commonly arise from extreme curvature and complex surface chemistry. Hair presents an especially difficult case due to its roughly 50-100 μm curvature radius, overlapping cuticle scales about 0.5 μm thick, and lipid-rich surface [36]. Using the WTT method, we transferred nanodiamond arrays onto hair in Figure 3c and obtained well-defined and continuous arrays in Figure S9c. This success may support on-hair detection of metabolic biomarkers such as cortisol and uric acid with surface-functionalized nanodiamonds. [37,38] The method's robustness was further confirmed by the successful transfer of arrays onto a silicone prosthetic hand with skin-like texture (Figure 3f), producing sharply defined patterns.

Collectively, these demonstrations across vastly different biointerfaces underscore the WTT method's efficacy in achieving conformal, large-area nanodiamond transfer on flexible and curved biointerfaces, significantly expanding the scope for in-situ biological detection and analysis.

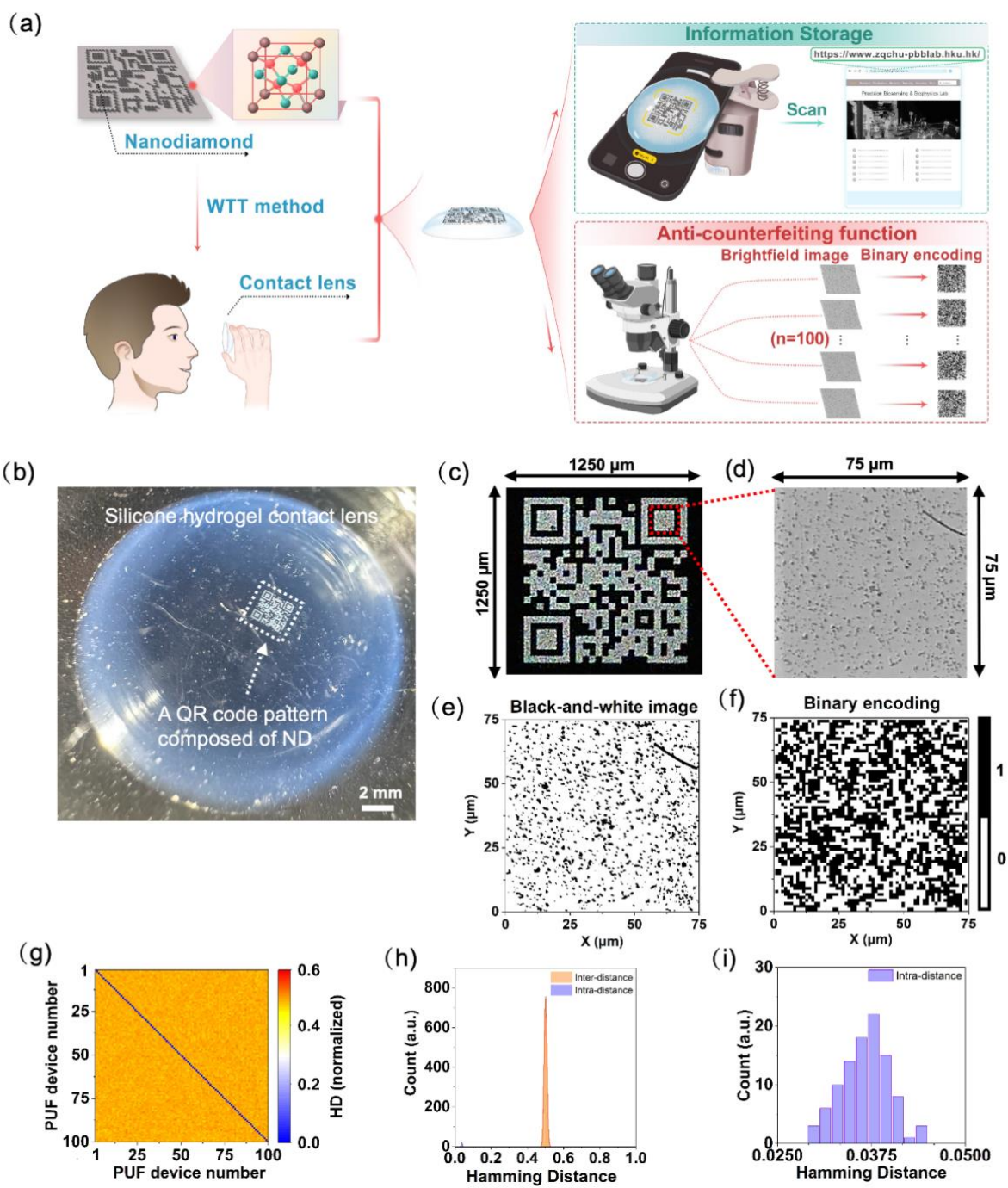

**Figure 4. Dual information recognition application of nanodiamond patterns on hydrogel contact lenses for anti-counterfeiting and data storage.** (a) Schematic illustration of the fabrication process and dual recognition method. (b) Photograph of a QR code pattern transferred onto a contact lens (c) Optical image of a 1250 μm QR code pattern transferred onto a contact lens. (d) Widefield optical image of one representative ND pattern from the 100 sampled devices; source data are available in Figure S12. (e) Corresponding binarized image (512×512 pixels) of the pattern in (d). (f) Downsampled binary-encoded image (64×64 pixels) derived from (e). (g) Pairwise comparison heatmap of 100 unique PUFs, showing an average

off-diagonal Hamming distance of 0.5049 (h) Histogram of inter- and intra-device Hamming distances for PUF responses. (i) Detailed histogram of intra-device Hamming distances (10 repeated measurements per device, mean value of 0.0366).

2.4 Dual information recognition application of nanodiamond patterns on flexible biomaterials

Wearable contact lenses present an ideal platform for next-generation identity authentication, capable of seamless integration with the human body for continuous and personalized security verification. [39,40] Analogous to iris recognition patterns, the contact lens surface itself can be engineered as a carrier for cryptographic information. [41] However, materializing this concept faces a dual challenge: first, the need for a high-security information medium that can be stably integrated onto the soft, hydrated, and curved surface of a contact lens; and second, the development of a manufacturing process that enables such patterning and integration without damaging the delicate biointerface. Nitrogen-vacancy center nanodiamonds, as an emerging quantum material, offer an exceptional platform for information encryption and anti-counterfeiting, owing to their superior photostability, programmable spin states and great promise for molecular detection and imaging [42-44] Here, by utilizing our WTT technique, we integrated nanodiamond arrays onto a silicone hydrogel contact lens, serving as a dual carrier for high-density information storage [45] and physical unclonable function (PUF) [46] security, thereby establishing a robust dual-identity verification system (**Figure 4**a).

First, we designed a NDs-based QR code pattern (1250 μm in size, Figure S10d) containing our laboratory's website URL and successfully transferred it onto a silicone hydrogel contact lens surface (Figure 4b). The QR code was reliably recognized using a portable microscope and smartphone, directing users to the intended webpage. The URL (31 characters) corresponds to 31 bytes or 248 bits, yielding a storage density of approximately 198 bit/mm² (based on a 1.250 mm² area); smartphone scanning took about 1 second with a read speed of approximately 248 bit/s, and the high recognition rate confirmed the reliability of the WTT method for information storage.

To investigate the anti-counterfeiting aspect of this dual identity verification, we leveraged the intrinsic stochasticity of nanodiamond ensembles, which has been established as a robust PUF due to the inherent randomness in particle morphology and distribution. [47,48] We thus evaluated whether our transferred ND arrays retained these critical PUF properties. The analysis focus on the QR code's alignment regions (150×150 μm squares at each corner for machine recognition). The ND particle distribution (75×75 μm area) within each region was imaged using bright-field mode on a wide-field microscope, with distribution optimized by adjusting ND particle concentration (Figure S11). Images were binarized and downsampled to 64×64 pixels to generate binary encoding. We fabricated and transferred 100 QR codes (Figure S12), capturing the ND distribution in the bottom-left alignment region of each. Analysis (Figures 4g–i) reveals: the histogram of inter- and intra-device Hamming distances (Figure 4h) shows an average inter-device Hamming distance of 0.5049, near the ideal value of 0.5, indicating strong distinguishability between devices; the intra-device Hamming distance (10 measurements per device, Figure 4i) averages 0.0366, below the 10% threshold, demonstrating excellent reproducibility; and the pairwise comparison heatmap (Figure 4g) confirms the uniqueness of each devices. These results collectively verify that the WTT method successfully preserves the high uniformity, uniqueness, and randomness of ND particle distributions on the soft contact lens, thereby validating the high fidelity and stability of our transfer process for implementing effective PUF-based security. This dual identity verification capability not only underscores the WTT method's superior performance on flexible bio-substrates but also opens new avenues for information security and anti-counterfeiting in smart biomaterial development.

## 3. Discussion

The development of transfer printing techniques has long been guided by a trade-off between conformability and precision. Solid carriers, such as elastomers [18] or adhesive tapes, [19] offer positional accuracy but fail on curved or soft surfaces; liquid carriers achieve conformability but sacrifice placement accuracy and incur buoyancy and hydrophobicity limitations. In the geometrically complex, topologically complex, moist, and fragile biological interfaces, traditional biocompatible solutions are almost impossible. The WTT method overcomes these limitations by introducing a transient, water-triggered solid-to-soft transition mechanism. This enables the conformal transfer of nanomaterial patterns onto soft, hydrated, and curvilinear biointerfaces, a feat that remains challenging for conventional approaches.

Our work introduces a distinct "active stress-assisted delamination model", which enabled by the unique swelling-dominated response of bilayer PVA tapes. Unlike the fracture-dominated release of solid stamps or the passive, diffusion-dominated dissolution of carriers like sugar-based reflows or composite tapes, this mechanism actively harnesses the material's hydration to perform mechanical work. The rapid, anisotropic swelling of the PVA backing (~500% swelling ratio increase) is not a mere physical change; it programmatically generates transient interfacial stresses. As quantified by TFM, this delivers a controlled compressive force to press the ND array against the substrate, while shear components proactively peel the carrier. Crucially, this entire process is triggered by water at room temperature, setting it apart from reflow-based methods that require thermal input (e.g., >80°C for sugar carriers), which poses risks of dehydration or thermal stress to biological tissues. We also performed stability test, the ND pattern on hydrogel contact lenses remained intact for over 1 month in room-temperature water (with detachment occurring after hydrogel collapse, Figure S13a), while patterns transferred onto PDMS showed no deformation or detachment after 10,000 stretch cycles at 10% strain (Figure S13b). By combining ambient-temperature processing with a self-driven, mechanically gentle release, the WTT method resolves the inherent trade-off between conformability and precision, specifically tailoring it for the demanding context of pristine bio-integration.

In real physiological microenvironments, ultra-low stiffness, sharp curvature, heterogeneous surface chemistry, and dynamic biofluids act synergistically to impose far more severe constraints on transfer efficiency than isolated model studies suggest. Ultra-soft tissues cause pronounced local deformation that drastically reduces contact area; high-curvature microstructures amplify non-linear stress and limit conformal contact; mixed charged/hydrated layers produce competing electrostatic and hydrogen-bonding interactions; and physiological fluids introduce ionic screening and wetting dynamics that weaken adhesion. As a result, the high transfer efficiencies (> 90%) routinely achieved ex vivo on simplified substrates (hydrogels, hair, prosthetics) are expected to decline significantly in vivo due to these coupled factors. Successful clinical translation therefore requires not only the inherently gentle water-triggered release mechanism, but also deliberate engineering of the biological interface to counteract the intrinsic variability of living microenvironments.

Looking forward, the evolution and application of the WTT platform can be advanced by addressing several key opportunities. First, while we have demonstrated high-efficiency transfer on various ex vivo biointerfaces, the long-term adhesion stability of transferred arrays under dynamic physiological conditions (e.g., in the presence of ions, proteins, pH fluctuations, and mechanical motion) remains to be systematically validated for in vivo applications. Developing next-generation water-soluble tapes with environmentally responsive properties (e.g., swelling kinetics and adhesion tunable by pH or ionic strength) will be crucial for enabling stable, long-term biosensing within living organisms. Second, to enhance process efficiency and versatility, exploring routes for direct nanopatterning on the carrier tape (e.g., via in situ self-assembly) could eliminate the initial pickup step from a rigid donor substrate, thereby minimizing material loss and simplifying the workflow. Finally, the generality of the WTT strategy is underscored by our successful transfer of other functional nanomaterials polymer microspheres (Figure S14), paving the way for its application in broader fields such as flexible electronics, cellular force mapping, and multimodal biosensing. When integrated with automated handling and 3D printing technologies, WTT has the potential to mature into a versatile, high-throughput manufacturing platform for integrating functional nanomaterials onto bio-interfaces.

In conclusion, we introduced a water-soluble tape transfer (WTT) strategy that effectively decouples the traditional trade-off between conformability and precision in micro-nanoparticle patterning. The method's success is based in a unique "Hydrate-soften-expand-self-peeling" mechanism of PVA-based tapes, which enables high-fidelity, residue-free transfer of nanodiamond arrays onto a wide range of soft, hydrated, and curvilinear biointerfaces. By systematically quantifying the influence of substrate stiffness, curvature, and chemistry, we have provided a rational framework for deploying this technology. The ability to implement dual-identity verification on a contact lens further highlights its potential for creating secure, smart biomedical devices. We anticipate that the WTT method, potentially augmented by automated handling and integration with 3D printing, will serve as a versatile platform for advancing bio-integrated quantum sensors, flexible electronics, and tissue engineering constructs.

## 4. Experimental Section

*Materials*: Silicon wafers were sourced from Jinghong Electronics Co., Ltd., China. The photoresist AZ5414E was obtained from Merck Performance Materials, US. 2.38% NMD developer was purchased from Tokyo Ohka Kogyo Co., Ltd., Japan. 3-Aminopropyltriethoxysilane (APTES, 99%), 3-(triethoxysilyl)propylsuccinic anhydride (TESP-SA, 99.5%), ethanol (99.95%), acetone (99.5%), isopropyl alcohol (IPA, 98%), deionized (DI) water and sodium chloride (NaCl, 99.5%) were purchased from Sigma-Aldrich, Germany. Water-soluble tapes were sourced from 3M Company, US (3M 5414 water-soluble wave solder tape), Shanghai Yongori Industrial Co., Ltd., China (Yongri water-soluble tape), and Tesa Tape Co., Ltd., China (Tesa 4446 water-soluble tape). Nanodiamonds (NDs, 150 nm, HPHT) were acquired from PolyQolor, China. Hydroxylated nanodiamond was purchased from FND Biotech, Inc., Taiwan. Methacrylated gelatin (GelMA, 90%) was obtained from Shanghai Aladdin Biochemical Technology Co., Ltd., China. Polydimethylsiloxane (PDMS) was purchased from Hangzhou Westru Technology Co., Ltd., China. Silicone hydrogel contact lenses were supplied by BenQ Materials Co., Ltd., Taiwan. Gold probes were obtained from Shenzhen Kaiyuan Electronics Co., Ltd., China.

*Preparation of Carboxylated NDs*: Carboxylated nanodiamonds were prepared using the salt-assisted air oxidation (SAAO) method. [49] Nanodiamonds (150 nm, HPHT) were mixed with NaCl in a 1:10 mass ratio, heated to 550 °C in a tube furnace under air flow for 4 h, washed thoroughly with deionized (DI) water to remove NaCl, and collected by centrifugation at 10,000 rpm for 10 min.

*Fabrication of ND Patterns on Silicon* Wafers: ND patterns were fabricated via surface functionalization, photolithography, and ND deposition. Silicon wafers were activated by $O_2$ plasma for 10 min (200 W) and then immediately immersed in 5% (v/v) APTES solution in ethanol for 1 h. The wafers were rinsed with acetone, isopropyl alcohol (IPA), and DI water, and dried at 70 °C for 30 min. AZ5414E photoresist was spin-coated at 4000 rpm for 40 s and soft-baked at 120 °C for 60 s. The template pattern was then defined by photolithography using a Microwriter ML3 or PM200 at a UV dose of 300 mJ/cm². The exposed resist was developed in 2.38% NMD solution for 60 s, rinsed with DI water, and dried under $N_2$ flow. The patterned wafer was immersed in 0.1 mg/mL ND solution for 30 min, rinsed with DI water, and dried at 70 °C for 30 min. Finally, the photoresist was removed by sequential rinsing with acetone, IPA, and DI water, followed by drying at 70 °C for 30 min.

*Characterization Methods:* Scanning electron microscopy (SEM) images were acquired using a Hitachi S-4800 SEM (Japan) or a LEO 1530 FEG SEM (Germany). Brightfield optical images were obtained using a VPPA53MET-FL microscope (Japan). Laser confocal microscopy was performed using a Zeiss LSM 980 confocal laser scanning microscope (Germany). X-ray photoelectron spectroscopy (XPS) measurements were conducted on a Thermo Scientific K-Alpha XPS system (US). Particle size and zeta potential were analyzed using a Malvern Zetasizer Nano ZS90 (UK).

*Characterization of Swelling Ratio of PVA Backing Layer* : The PVA backing layer was isolated by immersing strips of water-soluble tape (3M 5414) in DI water at room temperature for 30 min with gentle agitation to dissolve the adhesive layer. The isolated PVA film was retrieved, rinsed with fresh DI water, spread flat on aluminum foil, and dried overnight at 60 °C to constant weight. Dried PVA samples were weighed using an analytical balance (±0.001 g accuracy) and immersed in DI water at room temperature for 0, 2, 4, 7, 10, 20, 30, 60, and 180 s respectively. After removing the remaining surface water with filter papers, the PVA backing layer was immediately weighed to obtain the swollen mass $m_t$. The swelling ratio was calculated as:

$$SR_{mass} = \frac{m_t - m_0}{m_0} 100\%$$

where $m_t$ is the swollen mass of the PVA sample at each time point, and $m_0$ is the initial dry mass of the PVA sample before immersion. Measurements were performed in triplicate; results are reported as mean ± SD.

*Determination of Young's Modulus of Swollen PVA Backing Layer*: The Young's modulus of the swollen PVA backing layer was determined from rheological measurements. Isolated PVA backing films were stacked into bulky samples and immersed in deionized (DI) water at room temperature for 0, 2, 4, 7, 10, 20, 30, 60, and 180 s, resulting in rapid swelling and transformation into a hydrogel state. Rheological measurements were then performed on the swollen PVA samples using a rheometer (Thermo Scientific HAAKE, US) with a strain amplitude of 0.1% and a frequency of 1 Hz at 25 °C to obtain the storage modulus G′ and loss modulus G″. The Young's modulus E was calculated using the following equation: [50]

$$E \approx 3G'$$

Where E is Young's modulus, G′ is storage shear modulus. Measurements were performed in at least triplicate, and results are reported as mean ± standard deviation (SD).

*TFM Measurement during Tape Release*: The traction force field during the tape release process was determined using fluorescent nanodiamonds uniformly dispersed and adsorbed on glass substrates. This was achieved by coating the glass slides with an amine-modified PDMS layer. Specifically, glass substrates were first thoroughly cleaned in acetone, ethanol, and a UV-ozone cleaner. Subsequently, 100 μL of a PDMS base and curing agent mixture (60:1 mass ratio) was dispensed onto the surface. After spin-coating at 8000 rpm for 60 s, the glass substrates were cured overnight in a 60°C oven. To introduce amine functional groups on the surface for diamond adhesion, the PDMS-coated glass substrates were immersed in a 20:1 (v/v) ethanol: APTES solution for 20 min. After washing with ethanol and deionized water and drying under $N_2$ flow, the substrates were immersed in a 0.05 mg/mL fluorescent nanodiamond solution for 5 min to enable nanodiamond adsorption.

For the traction force field measurement, different types of water-soluble adhesive tapes were adhered to the fluorescent ND-coated PDMS surfaces. The samples were first imaged in the dry state using confocal laser scanning microscopy. After treating the tape with DI water for specific duration (20, 30, 40s), the post-swelling location of ND was imaged to calculation displacement. Displacement analysis was performed using ImageJ software. Image drift was first corrected using the "Align slices in stack" plugin. The "Particle Image Velocimetry (PIV)" plugin was then applied to compute displacement fields across the imaged area. Displacement vectors were reconstructed and visualized using the "Plot Particle Image Velocimetry" plugin to generate traction force maps.

*ND Pattern Transfer Process*: The water-soluble tape was cut to size, the Kraft paper liner was removed, and the adhesive side was pressed onto the patterned silicon wafer (using a PDMS block if necessary) to pick up the ND arrays. The tape carrying the patterns was then conformally attached to the target substrate. For small-area substrates, the tape was fixed adhesive-side up on a silicon wafer, and the target was rolled onto the patterns. The assembly was immersed in DI water for approximately 3 min; during this time the adhesive layer dissolved and the PVA backing swelled, generating stresses that detached it as a single piece. Gentle shaking accelerated the process. Once partial detachment occurred, the PVA backing could be removed with tweezers without affecting pattern integrity. For large-area transfers, multiple tape strips or batch processing was employed to ensure scalability and alignment.

*Preparation of GelMA Hydrogels*: Hydrogels with varying stiffness were prepared using methacrylated gelatin (GelMA, Shanghai Aladdin Biochemical Technology Co., Ltd.) as the base material. The stiffness was precisely controlled by adjusting the GelMA concentration. GelMA powder was dissolved in phosphate-buffered saline (PBS, pH 7.4) to prepare precursor solutions at concentrations of 15% (w/v), 10% (w/v), and 5% (w/v), corresponding to approximate Young's moduli of 18 kPa, 10 kPa, and 0.6 kPa, respectively. To improve adhesion of the hydrogels to the substrate, glass discs (10 mm diameter, 1 mm thickness) were surface treated prior to hydrogel fabrication. The discs were sequentially ultrasonicated in acetone, ethanol, and DI water for 10 min each. After drying, the discs were immersed in a 3-trimethoxysilylpropyl acrylate (Sigma-Aldrich) solution for 10 min, rinsed thoroughly with ethanol and DI water, and dried at 60 °C for 10 min. The GelMA precursor solutions (15%, 10%, and 5% w/v) were dispensed in 100 µL aliquots onto the treated glass discs to form a uniform droplet layer. The discs were then placed in a UV crosslinker (550 mW cm$^{-2}$, ZIGOO, China) for 2 min to initiate photo-crosslinking. After crosslinking, the hydrogels were immersed in PBS at 4 °C for 24 h before further use.

Mechanical characterization of the hydrogels was performed using atomic force microscopy (AFM, Bruker, Germany). A tipless cantilever functionalized with a 10 µm polystyrene bead was employed to acquire force–displacement curves on the hydrogel surface. The Young's modulus was extracted by fitting the retraction curves with the Hertz/Sneddon model. For each sample, three randomly selected regions on the hydrogel surface were probed, with each region indented in a 3 × 3 array of points.

*Preparation of PDMS Cylinder Models with Defined Curvatures* : PDMS cylinders were fabricated using the Sylgard 184 Silicone Elastomer Kit (Dow Corning). The PDMS base and curing agent were mixed at a 10:1 mass ratio and thoroughly stirred. The mixture was degassed in a vacuum chamber for 30 min to remove air bubbles introduced during mixing. The degassed PDMS was slowly poured into precision acrylic molds of varying diameters to minimize bubble formation. The filled molds were placed on a level surface and allowed to stand at room temperature (25 °C) for 30 min to release any residual microbubbles. The molds were then cured in an oven at 60 °C for 4 h. After cooling to room temperature, the PDMS cylinders were carefully demolded, with assistance from an air gun if necessary. This process yielded cylinders with diameters of 2 cm, 1.5 cm, 1 cm, 0.5 cm, and 0.25 cm, corresponding to curvatures of 1 cm$^{-1}$, 1.33 cm$^{-1}$, 2 cm$^{-1}$, 4 cm$^{-1}$, and 8 cm$^{-1}$, respectively.

*Surface Functionalization of PDMS*: For neutral -OH surfaces, PDMS was treated with O$_2$ plasma for 30 min. [51] For positively charged -NH$_2$ surfaces, plasma-activated PDMS was immediately immersed in 20:1 (v/v) ethanol: APTES solution for 30 min, rinsed with ethanol and DI water, and dried under N$_2$. [52] For negatively charged -COOH surfaces, plasma-activated PDMS was immersed in 20:1 (v/v) ethanol: TESP-SA solution for 20 min, followed by the same rinsing and drying procedure. [53]

*Readout and Digitization of the Diamond PUFs*: Diamond-based PUFs were readout using two setups to enable dual-identity verification (integrating data storage and unclonable anti-counterfeiting features): (i) a portable microscope equipped with a smartphone attachment for on-site rapid imaging and QR-code information extraction, and (ii) a home-built brightfield widefield microscope for high-resolution characterization and detailed PUF analysis. Image processing and quantitative analysis were performed using custom MATLAB scripts (R2024b). The following metrics were calculated to evaluate PUF performance: Uniformity was assessed based on the spatial distribution and binary consistency of the extracted PUF bit strings. Similarity index was computed to quantify the degree of pattern resemblance between different readout instances or devices. Hamming distance was used to measure the uniqueness and distinguishability among different PUF patterns.

*Statistical Analysis*: All data are presented as mean ± standard deviation (SD) except where specifically indicated. Statistical comparisons between groups were assessed using one-way ANOVA followed by Tukey's multiple comparison test. A p-value less than 0.05 was considered statistically significant (*$p < 0.05$; **$p < 0.01$; ***$p < 0.001$; ****$p < 0.0001$). Analyses were conducted using GraphPad Prism version 8.

**Acknowledgements:** Z.Q.C. acknowledges the financial support from the National Natural Science Foundation of China (NSFC) and the Research Grants Council (RGC) of the Hong Kong Joint Research Scheme (Project No. N_HKU750/23), HKU seed fund, and the Shenzhen-Hong Kong-Macau Technology Research Programme (Category C project, no. SGDX20230821091501008). Y.L. acknowledges support from the Research Grants Council of Hong Kong under the General Research Fund (Grant no. 17210520) and the National Natural Science Foundation of China (Grant no. 12272332).

**Author contributions:** Z.Q.C., Y.H. and L.Y.Z. conceived the project and designed the experiments. L.Y.Z. performed the measurements, data analysis and wrote the required programs under the guidance of Y.H. L.Y.Z. and Y.H. and drafted the manuscript. L.Z.W. assisted with assisted with anti-counterfeiting data collection and analysis. X.H.H. assisted with Young's modulus testing. N.T.Y. and J.X.J. assisted with sample preparation. Z.M.S. and Z. L. X. assisted with method research. Z.Q.C. Y.H. and Y.L supervised the project. All authors discussed the results and contributed to the writing of the manuscript.

**Competing interests:** The authors declare no competing interests.

**Data and materials availability:** All data are available in the main text or the supplementary materials.

Received: ((will be filled in by the editorial staff))

Revised: ((will be filled in by the editorial staff))

Published online: ((will be filled in by the editorial staff))

# References

[1] Y. Wu, and T. Weil, "Recent Developments of Nanodiamond Quantum Sensors for Biological Applications," *Advanced Science* 9, no. 19 (2022): 2200059, https://doi.org/10.1002/advs.202200059

[2] N. Savage, "Quantum Diamond Sensors," *Nature* 591, S37 (2021): https://doi.org/10.1038/d41586-021-00742-4

[3] Z. Du, M. Gupta, F. Xu, K. Zhang, J. Zhang, Y. Zhou, Y. Liu, Z. Wang, J. Wrachtrup, N.Wong, C.Li, and Z. Chu "Widefield Diamond Quantum Sensing with Neuromorphic Vision Sensors," *Advanced Science* 11 , no. 2 (2024): 2470013, https://doi.org/10.1002/advs.202304355

[4] Q. S. Ahmad, W. W. Hsiao, L. Hussain, H. Aman, T. Le, and M. Rafique, "Recent Development of Fluorescent Nanodiamonds for Optical Biosensing and Disease Diagnosis," *Biosensors* 12, no. 12 (2022): 1181, https://doi.org/10.3390/bios12121181

[5] Y. Feng, Q. Zhao, Y. Shi, G. Gao, and J. Zhi, "Recent Applications of Fluorescent Nanodiamonds Containing Nitrogen-Vacancy Centers in Biosensing," *Functional Diamond* 2, no.1 (2022): 192-203, https://doi.org/10.1080/26941112.2022.2153233

[6] J. Rovny, Z. Yuan, M. Fitzpatrick, A. I. Abdalla, L. Futamura, C. Fox, M. C. Cambria, S. Kolkowitz, and N. P. de Leon, "Nanoscale Covariance Magnetometry with Diamond Quantum Sensors," *Science* 378, no.6626 (2022): 1301-1305, https://doi.org/10.1126/science.ade9858

[7] D. R Glenn, K. Lee, H. Park, R. Weissleder, A. Yacoby, M. D Lukin, H. Lee, R. L Walsworth, and Colin B Connolly, "Single-Cell Magnetic Imaging Using a Quantum Diamond Microscope," *Nature Methods* 12, no.8 (2015): 736-738, https://doi.org/10.1038/nmeth.3449

[8] J. Lazovic, E. Goering, A. Wild, P. Schützendübe, A. Shiva, J. Löffler, G. Winter, and M. Sitti, "Nanodiamond-Enhanced Magnetic Resonance Imaging," *Advanced Materials* 36, no. 11 (2024): 2310109, https://doi.org/10.1002/adma.202310109

[9] W. W. Hsiao, Y. Y. Hui, P. Tsai, and H. Chang, "Fluorescent Nanodiamond: A Versatile Tool for Long-Term Cell Tracking, Super-Resolution Imaging, and Nanoscale Temperature Sensing," *Accounts of Chemical Research* 49, no.3 (2016): 400-407, https://doi.org/10.1021/acs.accounts.5b00484

[10] T. Plakhotnik, M. Doherty, J. Cole, R. Chapman, and N. B. Manson, "All-Optical Thermometry and Thermal Properties of the Optically Detected Spin Resonances of the NV− Center in Nanodiamond," *Nano Letters* 14, no.9 (2014): 4989-4996, https://doi.org/10.1021/nl501841d

[11] R. Dou, G. Zhu, W. Leong, X. Feng, Z. Li, C. Lin, S. Wang, and Q. Li, "In Operando Nanothermometry by Nanodiamond Based Temperature Sensing," *Carbon* 203, (2023): 534-541, https://doi.org/10.1016/j.carbon.2022.11.075

[12] G. Petrini, G. Tomagra, E. Bernardi, E. Moreva, P. Traina, A. Marcantoni, F. Picollo, K. Kvaková, P. Cígler, I. P. Degiovanni, V. Carabelli, and M. Genovese, "Nanodiamond-Quantum Sensors Reveal Temperature Variation Associated to Hippocampal Neurons Firing," *Advanced Science* 9, no. 28 (2022): 2202014, https://doi.org/10.1002/advs.202202014

[13] L. P. McGuinness, Y. Yan, A. Stacey, D. A. Simpson, L. T. Hall, D. Maclaurin, S. Prawer, P. Mulvaney, J. Wrachtrup, F. Caruso, R. E. Scholten, and L. C. L. Hollenberg, "Quantum Measurement and Orientation Tracking of Fluorescent Nanodiamonds Inside Living Cells," *Nature Nanotechnology* 6, (2011): 358-363, https://doi.org/10.1038/nnano.2011.64

[14] Z. Qin, Z. Wang, F. Kong, J. Su, Z. Huang, P. Zhao, S. Chen, Q. Zhang, F. Shi, and J. Du, "In Situ Electron Paramagnetic Resonance Spectroscopy Using Single Nanodiamond Sensors," *Nature Communications* 14, (2023): 6278. https://doi.org/10.1038/s41467-023-41903-5

[15] T. Fujisaku, R. Tanabe, S. Onoda, R. Kubota, T. F. Segawa, F. T. So, T. Ohshima, I. Hamachi, M. Shirakawa, and R. Igarashi, "pH Nanosensor Using Electronic Spins in Diamond," *ACS Nano* 13, no. 10, (2019): 11726-11732, https://doi.org/10.1021/acsnano.9b05342

[16] L. Wang, Y. Hou, T. Zhang, X. Wei, Y. Zhou, D. Lei, Q. Wei, Y. Lin, and Z. Chu, "All-Optical Modulation of Single Defects in Nanodiamonds: Revealing Rotational and Translational Motions in Cell Traction Force Fields," *Nano Letters* 22, no.18 (2022): 7714–7723, https://doi.org/10.1021/acs.nanolett.2c02232

[17] H. Huang, E. Pierstorff, E. Osawa, and D. Ho, "Active Nanodiamond Hydrogels for Chemotherapeutic Delivery," *Nano Letters* 7, no.11 (2007): 3305-3314, https://doi.org/10.1021/nl071521o

[18] M. A. Meitl, Z. Zhu, V. Kumar, K. J. Lee, X. Feng, Y. Y. Huang, I. Adesida, R. G. Nuzzo, and J. A. Rogers, "Transfer Printing by Kinetic Control of Adhesion to an Elastomeric Stamp," *Nature Materials* 5, (2006): 33-38, https://doi.org/10.1038/nmat1532

[19] Z. Yan, T. Pan, M. Xue, C. Chen, Y. Cui, G. Yao, L. Huang, F. Liao, W. Jing, H. Zhang, M. Gao, D. Guo, Y. Xia, and Y. Lin, "Micro-/Nanostructured Mechanical and Chemical Sensors Enabled by Transfer Printing," *Advanced Science* 4, no.11 (2017): 1700251, https://doi.org/10.1002/advs.201700251

[20] G. F. Schneider, V. E. Calado, H. Zandbergen, L. M. K. Vandersypen, and C. Dekker, "Wedging Transfer of Nanostructures," *Nano Letters* 10, no.5 (2010): 1912-1916, https://doi.org/10.1021/nl1008037

[21] M. Aghajamali, I T. Cheong, and J. G. C. Veinot, "Water-Assisted Transfer Patterning of Nanomaterials," *Langmuir* 34, no.32 (2018): 9418-9423, https://doi.org/10.1021/acs.langmuir.8b00694

[22] G. Zabow, "Reflow Transfer for Conformal Three-Dimensional Microprinting," *Science* 378, no. 6622 (2022): 894-898, https://doi.org/10.1126/science.add7023

[23] J. Weickenmeier, R. D. Rooij, S. Budday, P. Steinmann, T. C. Ovaert, and E. Kuhl, "Brain Stiffness Increases with Myelin Content," *Acta Biomaterialia* 42, no.15 (2016): 265-272, https://doi.org/10.1016/j.actbio.2016.07.040

[24] D. L. Rorbinson, M. E. Kersh, N. C. Walsh, D. C. Ackland, R. N. Steiger, and M. G. Pandy, "Mechanical Properties of Normal and Osteoarthritic Human Articular Cartilage," *Journal of the Mechanical Behavior of Biomedical Materials* 61, (2016): 96-109 , https://doi.org/10.1016/j.jmbbm.2016.01.015

[25] B. Kasemo, "Biological Surface Science," *Surface Science* 500, no. 1-3 (2002): 656-677, https://doi.org/10.1016/S0039-6028(01)01809-X

[26] A. E. Nel, L. Mädler, D. Velegol, T. Xia, E. M. V. Hoek, P. Somasundaran, F. Klaessig, V. Castranova, and M. Thompson, "Understanding Biophysicochemical Interactions at The Nano–Bio Interface," *Nature materials* 8, (2009): 543-557, https://doi.org/10.1038/nmat2442

[27] T. Ho, C. Chang, H. Chan, T. Chung, C. Shu, K. Chuang, T. Duh, M. Yang, and Y. Tyan, "Hydrogels: Properties and Applications in Biomedicine," *Molecules* 27, no. 9 (2022): 2902, https://doi.org/10.3390/molecules27092902

[28] A. B. Mathur, A. M. Collinsworth, W. M. Reichert, W. E. Kraus, and G. A. Truskey, "Endothelial, Cardiac and Skeletal Muscle Exhibit Different Viscous and Elastic Properties as Determined by Atomic Force Microscopy," *Journal of Biomechanics* 34, no. 12 (2001): 1545-1553, https://doi.org/10.1016/S0021-9290(01)00149-X

[29] W. Yeh, P. Li, Y. Jeng, H. Hsu, P. Kuo, M. Li, P. Yang, and P. H. Lee, "Elastic Modulus Measurements of Human Liver and Correlation with Pathology," *Ultrasound in Medicine & Biology* 28, no.4 (2002): 467-474, https://doi.org/10.1016/S0301-5629(02)00489-1

[30] L. Pastewka, and M. O. Robbins, "Contact between rough surfaces and a criterion for macroscopic adhesion," *Proceedings of the National Academy of Sciences* 111, no.9 (2014): 3298-3303, https://doi.org/10.1073/pnas.1320846111

[31] K. R. Shull, "Contact Mechanics and the Adhesion of Soft Solids," *Materials Science and Engineering: R: Reports* 36, no. 1 (2002): 1-45, https://doi.org/10.1016/S0927-796X(01)00039-0

[32] N. Efron, N. A. Brennan, R. L. Chalmers, L. Jones, C. Lau, P. B. Morgan, J. J. Nichols, L. B. Szczotka-Flynn, and M. D. Willcox, "Thirty Years of 'Quiet Eye' with Etafilcon a Contact Lenses," *Contact Lens and Anterior Eye* 43, no. 3 (2020): 285-297, https://doi.org/10.1016/j.clae.2020.03.015

[33] J. Zhang, Z. Cheng, P. Li, and B. Tian, "Materials and Device Strategies to Enhance Spatiotemporal Resolution in Bioelectronics," *Nature Reviews Materials* 10, (2025): 425-448, https://doi.org/10.1038/s41578-025-00798-y

[34] P. Fan, Y. Liu, Y. Pan, Y. Ying, and J. Ping, "Three-Dimensional Micro- and Nanomanufacturing Techniques for High-Fidelity Wearable Bioelectronics," *Nature Reviews Electrical Engineering* 2, (2025): 390-406, https://doi.org/10.1038/s44287-025-00174-6

[35] J. Wu, H. Liu, W. Chen, B. Ma, and H. Ju, "Device Integration of Electrochemical Biosensors," *Nature Reviews Bioengineering* 1, (2023): 346-360, https://doi.org/10.1038/s44222-023-00032-w

[36] B. Bhushan 2010. "Morphology and Structure of Human Hair" in *Biophysics of Human Hair: Structural, Nanomechanical, and Nanotribological Studies*. Springer, Berlin: Heidelberg. ISBN 978-3-642-15900-8.

[37] M. B. A. Olia, P. S. Donnelly, L. C. L. Hollenberg, P. Mulvaney, and D. A. Simpson, "Advances in the Surface Functionalization of Nanodiamonds for Biological Applications: A Review," *ACS Applied Nano Materials* 4, no. 10 (2021): 9985-10005, https://doi.org/10.1021/acsanm.1c02698

[38] D. H. Jariwala, D. Parel and S. Wairkar, "Surface Functionalization of Nanodiamonds for Biomedical Applications," *Materials Science and Engineering: C* 113, (2020): 110996, https://doi.org/10.1016/j.msec.2020.110996

[39] J. Kim, M. Kim, M. Lee, K. Kim, S. Ji, Y. Kim, J. Park, K. Na, K. Bae, H. K. Kim, F. Bien, C. Y. Lee, and J. Park, "Wearable Smart Sensor Systems Integrated on Soft Contact Lenses for Wireless Ocular Diagnostics," *Nature Communications* 8, (2017): 14997, https://doi.org/10.1038/ncomms14997

[40] L. M. Shaker, A. AI-Amiery, M. S. Takriff, W. N. R. W. Isahak, A. S. Mahdi and W. K. AI-Azzawi, "The Future of Vision: A Review of Electronic Contact Lenses Technology," *ACS Photonics* 10, no. 6 (2023): 1671-1686, https://doi.org/10.1021/acsphotonics.3c00523

[41] Y. Xia, M. Khamis, F. A. Fernandez, H. Heidari, H. Butt, and Z. Ahmed, "State-of-the-Art in Smart Contact Lenses for Human–Machine Interaction," *IEEE Transactions on Human-Machine Systems* 53, no.1 (2022): 187-200, https://doi.org/10.1109/THMS.2022.3224683

[42] T. Zhang, G. Pramanik, K. Zhang, M. Gulka, L. Wang, J. Jing, F. Xu, Z. Li, Q. Wei, P. Cigler and Z. Chu, "Toward Quantitative Bio-Sensing with Nitrogen–Vacancy Center in Diamond," *ACS Sensors* 6, no. 6 (2021): 2077–2107, https://doi.org/10.1021/acssensors.1c00415

[43] J. Zhang, L. Ma, Y. Hou, H. Ouyang, H. Hong, K. Kim, H. Kang, Z. Chu, "Nanodiamond-Based Sensing: A Revolution for Biosensors in Capturing Elusive Bio-Signals in Living Cells," *Advanced Drug Delivery Reviews* 221, (2025): 115590, https://doi.org/10.1016/j.addr.2025.115590

[44] F. Xu, S. Zhang, L. Ma, Y. Hou, J. Li, A. Denisenko, Z. Li, J. Spatz, J. Wrachtrup, H. Lei, Y. Cao, Q. Wei, and Z. Chu, "Quantum-Enhanced Diamond Molecular Tension Microscopy for Quantifying Cellular Forces," *Science Advances* 10, no. 4 (2024): eadi5300, https://doi.org/10.1126/sciadv.adi5300

[45] Y. Sun, X. Le, S. Zhou, and T. Chen, "Recent Progress in Smart Polymeric Gel-Based Information Storage for Anti-Counterfeiting," *Advanced Materials* 34, no. 41 (2022): 2201262, https://doi.org/10.1002/adma.202201262

[46] Y. Gao, S. F. AI-Sarawi, and D. Abbott, "Physical Unclonable Functions," *Nature Electronics* 3, (2020): 81-91, https://doi.org/10.1038/s41928-020-0372-5

[47] T. Zhang, L. Wang, J. Wang, Z. Wang, M. Gupta, X. Guo, Y. Zhu, Y. C. Yiu, T. K. C. Hui, Y. Zhou, C. Li, D. Lei, K. H. Li, X. Wang, Q. Wang, L. Shao and Z. Chu, "Multimodal Dynamic and Unclonable Anti-Counterfeiting Using Robust Diamond Microparticles on Heterogeneous Substrate," *Nature Communications* 14, (2023): 2507, https://doi.org/10.1038/s41467-023-38178-1

[48] L. Wang, X. Yu, T. Zhang, Y. Hou, D. Lei, X. Qi and Z. Chu, "High-Dimensional Anticounterfeiting Nanodiamonds Authenticated with Deep Metric Learning," *Nature Communications* 15 (2024): 10602, https://doi.org/10.1038/s41467-024-55014-2

[49] T. Zhang, L. Ma, L. Wang, F. Xu, Q. Wei, W. Wang, Y. Lin, and Z. Chu, "Scalable fabrication of clean nanodiamonds via salt-assisted air oxidation: implications for sensing and imaging," *ACS Applied Nano Materials* 4, no.9 (2021): 9223-9230, https://doi.org/10.1021/acsanm.1c01751

[50] U. Chippada, B. Yurke, and N. A. Langrana, "Simultaneous Determination of Young's Modulus, Shear Modulus, and Poisson's Ratio of Soft Hydrogels," *Journal of Materials Research* 25, no.3 (2010): 545-555, https://doi.org/10.1557/JMR.2010.0067

[51] S. H. Tan, N. Nguyen, Y. C. Chua, and T. G. Kang, "Oxygen plasma treatment for reducing hydrophobicity of a sealed polydimethylsiloxane microchannel," *Biomicrofluidics* 4, no.3 (2010): 032204, https://doi.org/10.1063/1.3466882

[52] A. AI-Ali, W. Waheed, F. Dawaymeh, N. Alamoodi, and A. Alazzam, "A Surface Treatment Method for Improving the Attachment of PDMS: Acoustofluidics as A Case Study," *Scientific Reports* 13, (2023): 18141, https://doi.org/10.1038/s41598-023-45429-0

[53] G. Liu, M. Gao, W. Chen, X. Hu, L. Song, B. Liu, and Y. Zhao, "pH-Modulated Ion-Current Rectification in A Cysteine-Functionalized Glass Nanopipette," *Electrochemistry Communications* 97, (2018): 6-10, https://doi.org/10.1016/j.elecom.2018.09.017

ToC figure

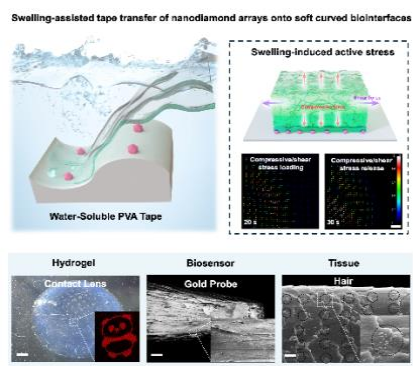

This work introduces a swelling-induced, stress-assisted water-soluble PVA tape strategy to transfer-print nanodiamond quantum-sensor arrays onto soft, curved biological interfaces. The room-temperature, water-triggered process achieves >98% fidelity and residue-free integration, enabling conformal quantum sensing on contact lenses, neural probes, and hair for bio-integrated diagnostics and anti-counterfeiting.

# Supporting Information

**Experimental Details, Stability Testing, and Transfer Efficiency Analysis for Nanodiamond Patterning via the WTT Method**

*Luyao Zhang[1], Lingzhi Wang[1], Xinhao Hu[1], Yip Tai Nam[1], Mingzhe Sun[2], Jixiang Jing[1], Lizhi Xu[2], Yuan Lin[2\*], Yong Hou[1\*], Zhiqin Chu[1,2,3\*].*

1.     Department of Electrical and Electronic Engineering, The University of Hong Kong, Pok Fu Lam, Hong Kong.

2.     Department of Mechanical Engineering, The University of Hong Kong, Pok Fu Lam, Hong Kong.

3.     School of Biomedical Engineering, The University of Hong Kong, Pok Fu Lam, Hong Kong.

\* Corresponding authors:

Prof. Dr. Yuan Lin, Email: ylin@hku.hk

Dr. Yong Hou, Email: houyong@eee.hku.hk

Prof. Dr. Zhiqin Chu, Email: zqchu@eee.hku.hk

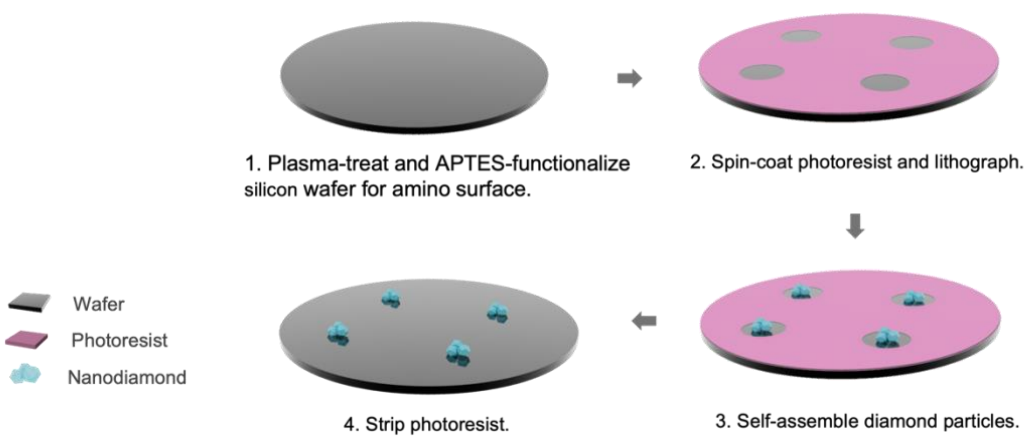

**Figure S1. Fabrication process for patterning diamond nanoparticles on a silicon wafer**:
1. Plasma treatment and APTES functionalization to create an amino-terminated surface. 2.

Spin-coating photoresist followed by lithography. 3. Self-assembly of diamond nanoparticles, and 4. Photoresist removal.

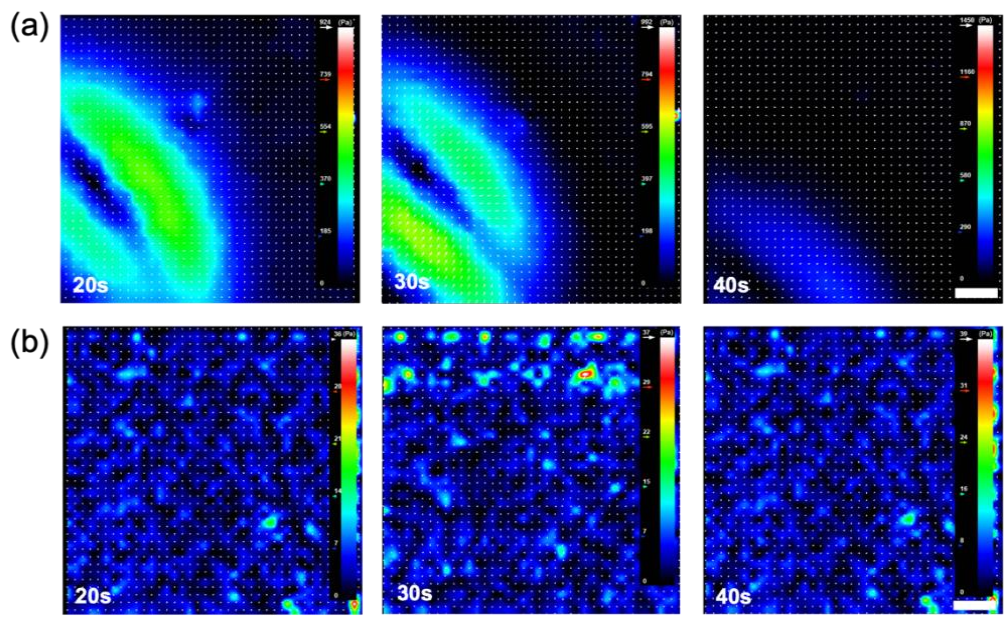

**Figure S2**. **Traction force maps (TFM) revealing the force dynamics at the tape-substrate interface during tape Swelling and dissolution**. (a) traction force and displacement vector map for the swelling-assisted tape, (b) traction force and displacement vector map for the dissolution-based tape.

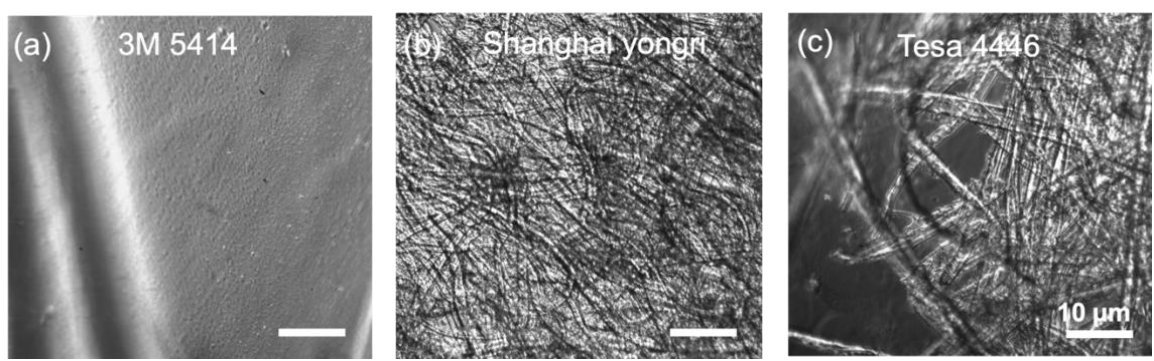

**Figure S3. Water-soluble tapes from various brands and compositions.** (a) 3M 5414 water-soluble wave solder tape, (b) Shanghai Yongri water-soluble tape (c) Tesa 4446 water-soluble tape

| Mechanism | Product name | Composition | Thickness | Degradation time | adhesion |
|---|---|---|---|---|---|
| Swelling | 3M 5414 Water-Soluble Wave Solder Tape | PVA | 53um | 3min | Strong |

| Dissolution | Shanghai Yongri Water-soluble tape | Cellulose fiber | 50-100um | 30min | Normal |
| Dissolution | Tesa 4446 Water-Soluble Tape | Cellulose fiber | 115um | 5min | Strong |

**Table S1 Different water-soluble tapes comparison**

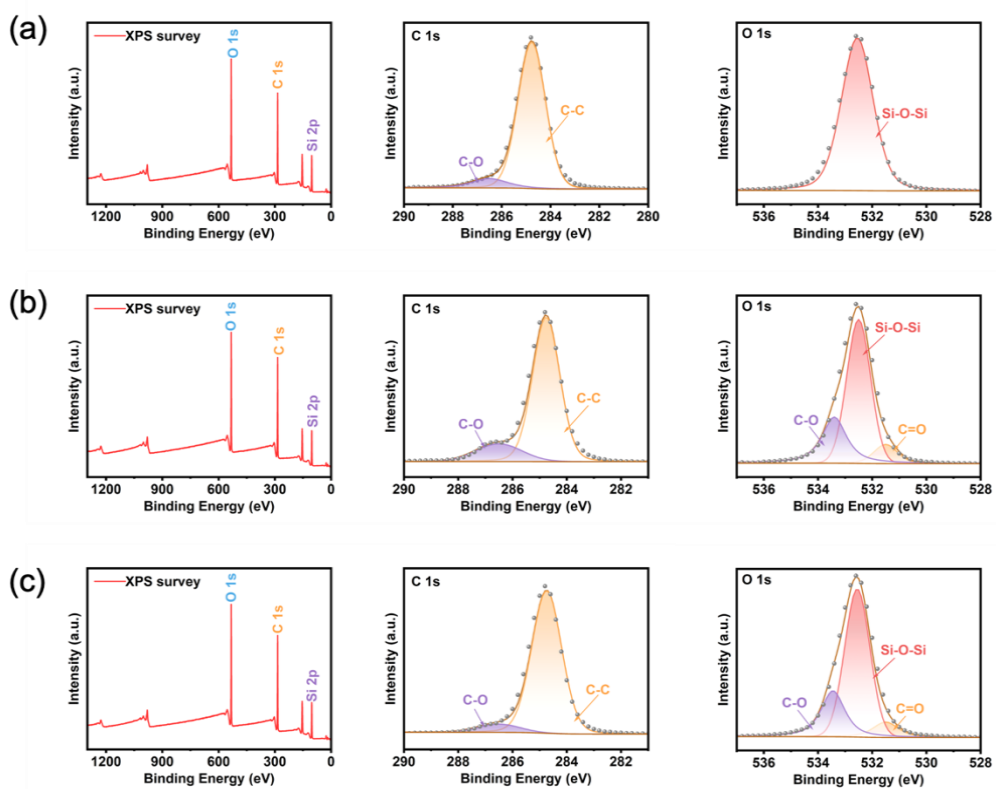

**Figure S4. X-ray photoelectron spectroscopy (XPS) analysis of tape residues on PDMS substrates after different release processes.** (a) Residue characterization after the swelling-assisted release process using 3M 5414 tape. (b,c) Residue characterization after the dissolution-based release process using Tesa (b) and Yongri (c) tapes.

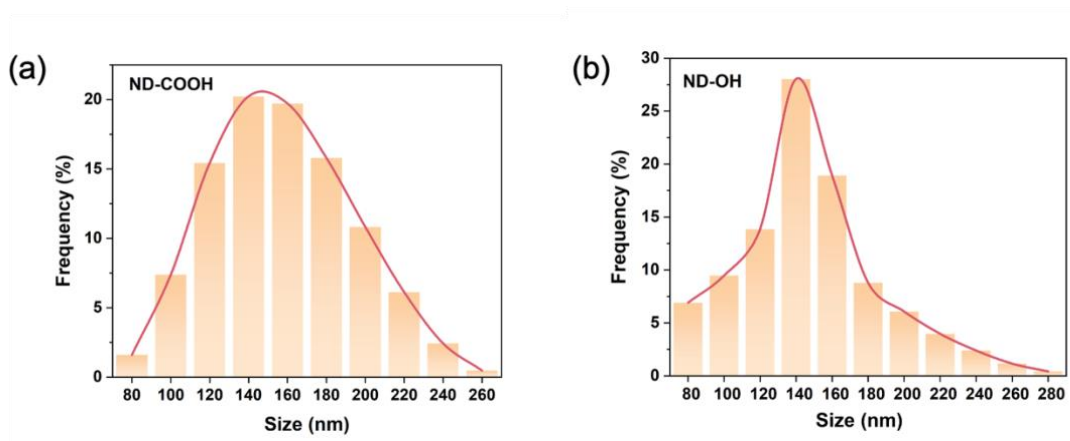

**Figure S5. Size frequency distribution of functionalized NDs**. (a) Size frequency distribution of carboxyl-functionalized nanodiamonds (ND-COOH). (b) Size frequency distribution of hydroxyl-functionalized nanodiamonds (ND-OH).

| Sample | ND-COOH | ND-OH |
|---|---|---|
| Zeta potential (mV) | -45.4 | -10.5 |

**Table S2. The zeta potentials of NDs with different functional groups**

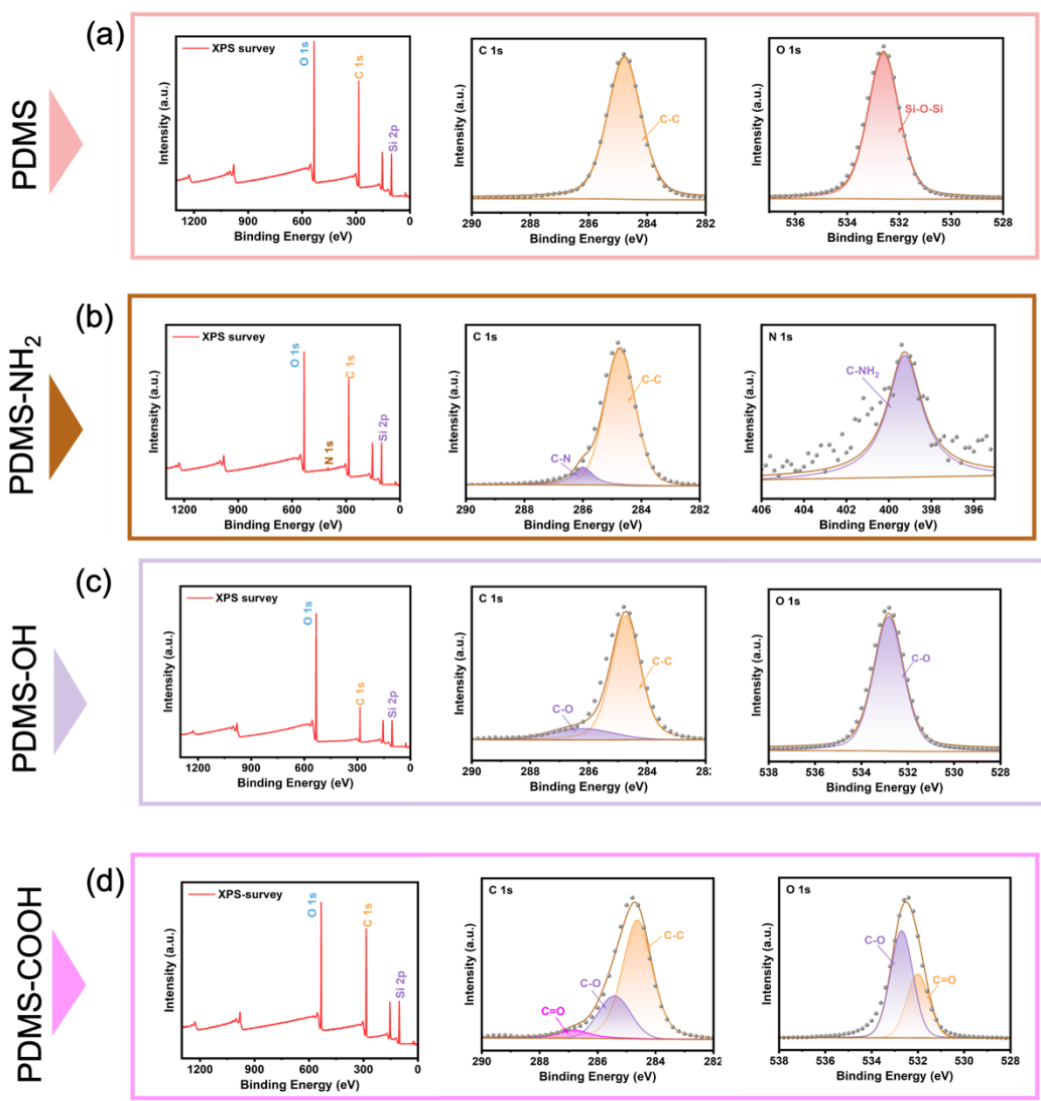

**Figure S6. XPS Analysis of Functionalized PDMS Substrates.**

![WILEY-VCH]

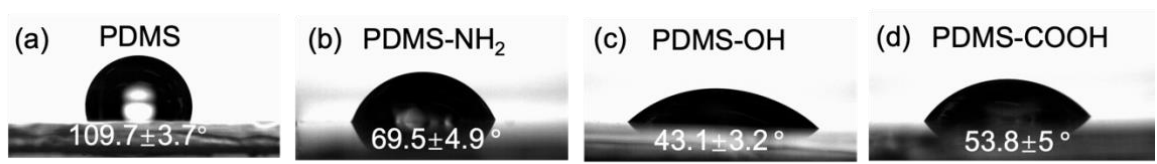

**Figure S7. Contact Angle (CA) measurements of Functionalized PDMS Substrates.**

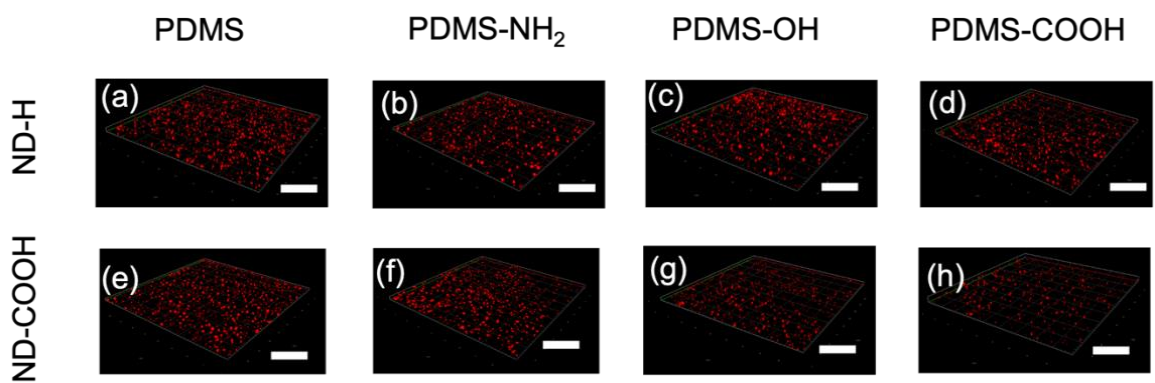

**Figure S8. Confocal fluorescence images of ND-OH and ND-COOH transfer-printed onto PDMS with different functional groups over a 40 µm × 40 µm area.**

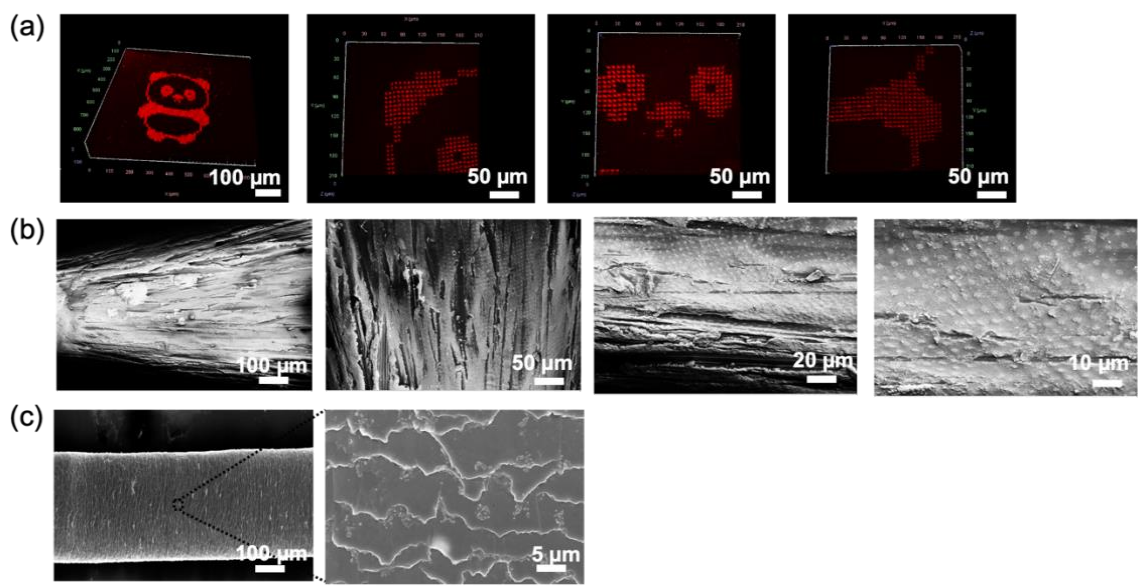

**Figure S9**. **Additional views from Figure 3**. (a) Silicone hydrogel contact lens with panda pattern and enlarged detail. (b) Gold probe bearing a circular ND array (c) Hair bearing a circular ND array (4 µm diameter circles, 10 µm period).

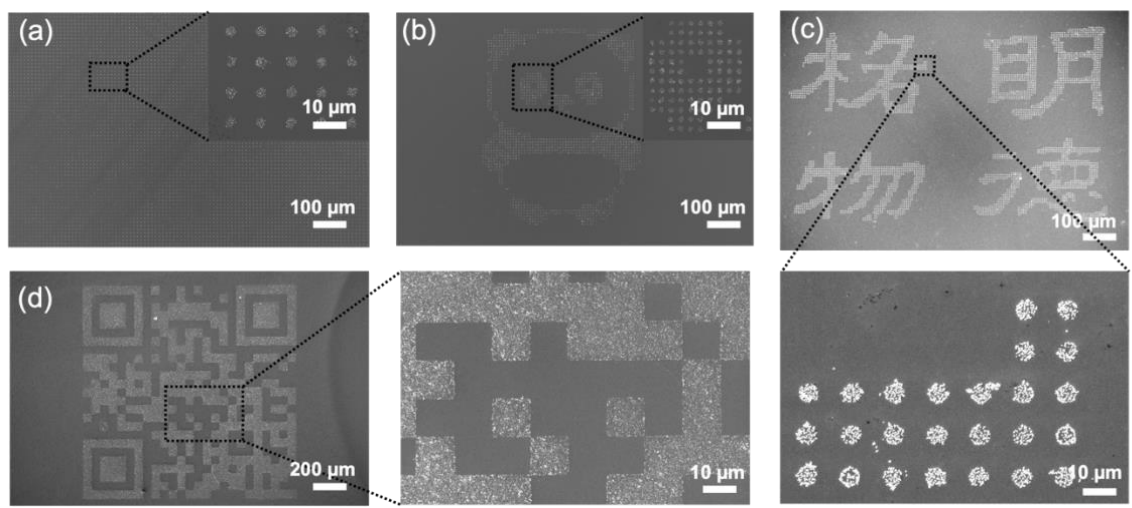

**Figure S10. The ND pattern on silicon wafer before transfer**.

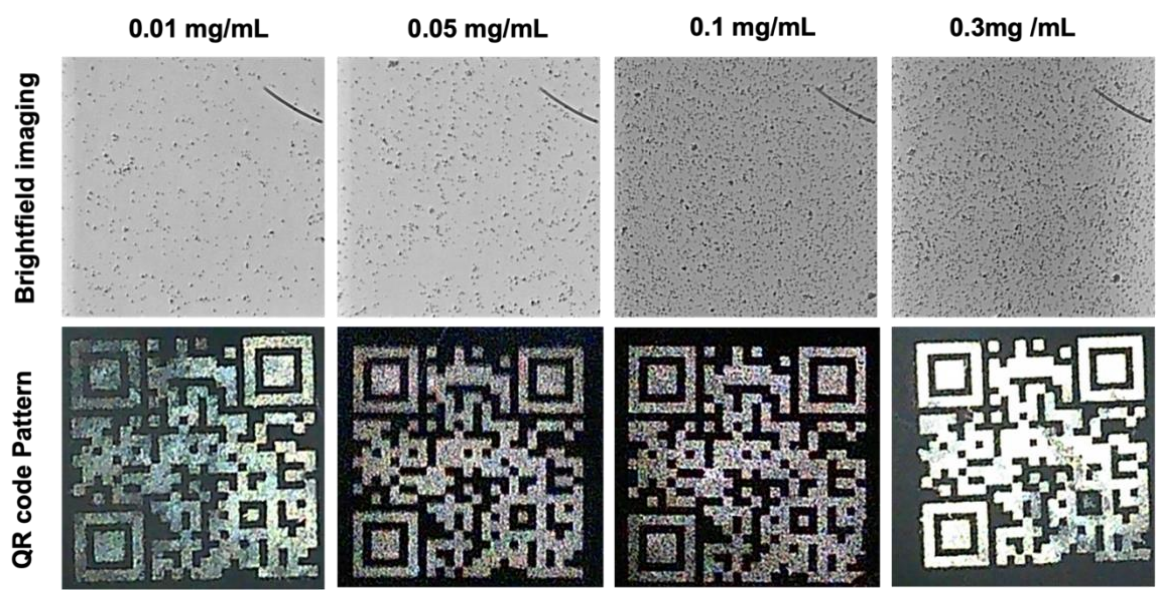

**Figure S11**. QR code patterns transferred onto contact lenses after incubation in diamond nanoparticle solutions at different, shown as partial views under widefield microscopy and full views under portable microscopy.

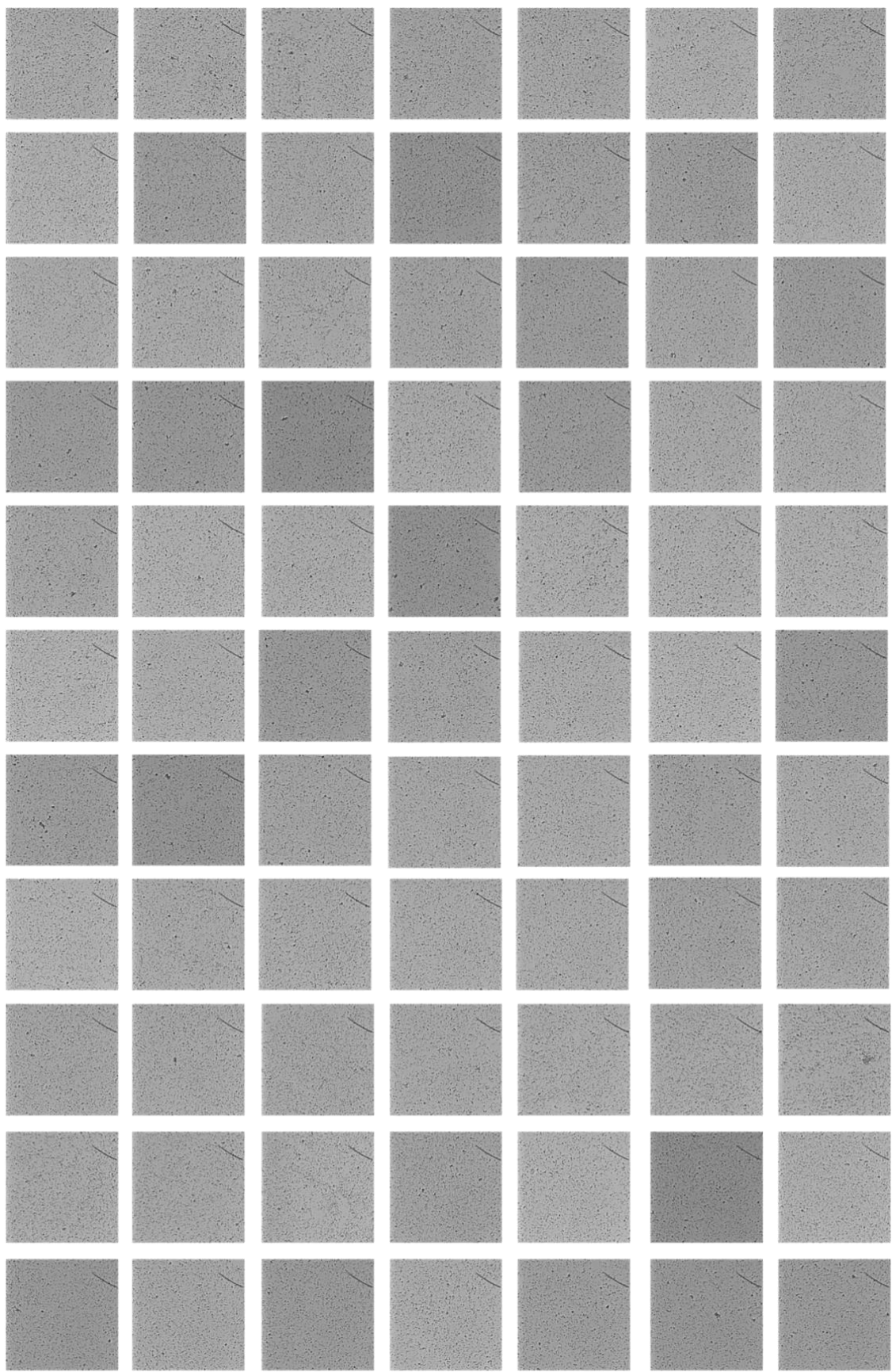

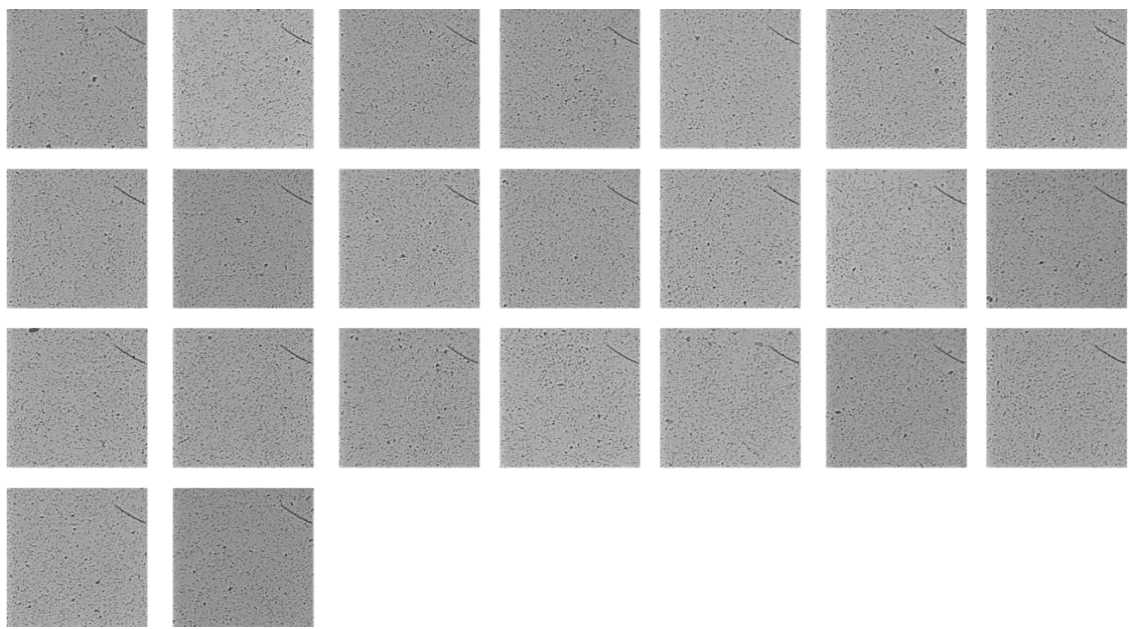

**Figure S12**. Brightfield widefield microscopy images of 100 contact lens-based physical unclonable function (PUF) devices fabricated for anti-counterfeiting demonstration.

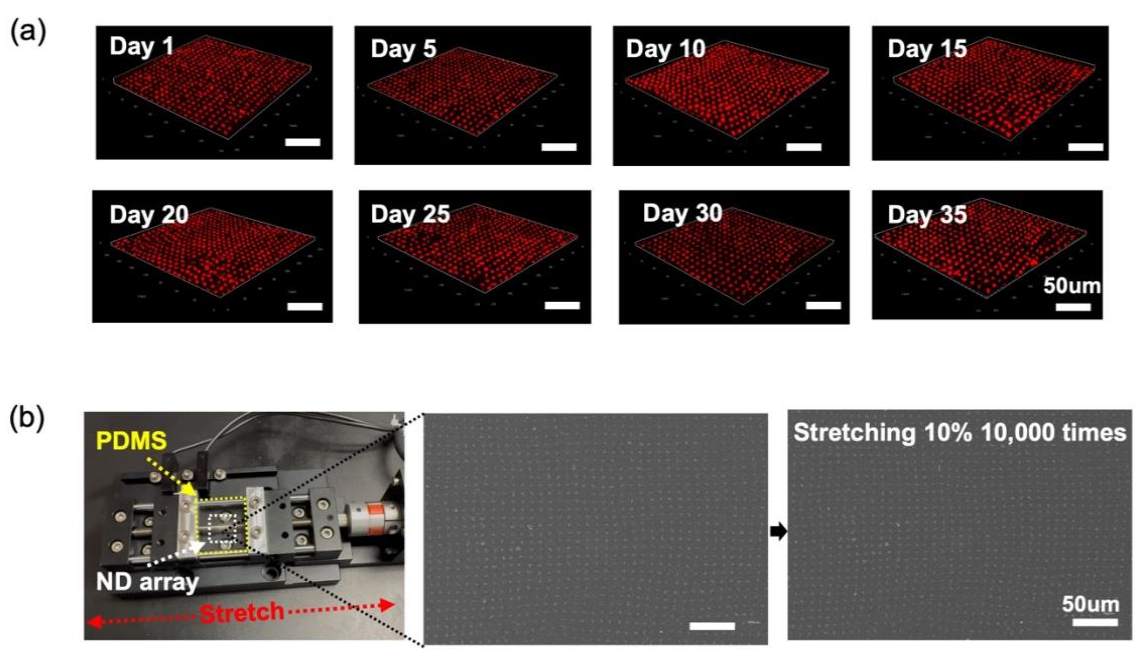

**Figure S13. Stability evaluation of transferred ND patterns under aqueous and mechanical stress conditions.** (a) ND patterns on hydrogel immersed in aqueous media, showing stable retention over 35 days. (b) SEM images of ND-patterned PDMS before and after 10000 cycles of 10% tensile strain: (left) schematic of the tensile stretching experiment, (center) SEM image before stretching, (right) SEM image after stretching.

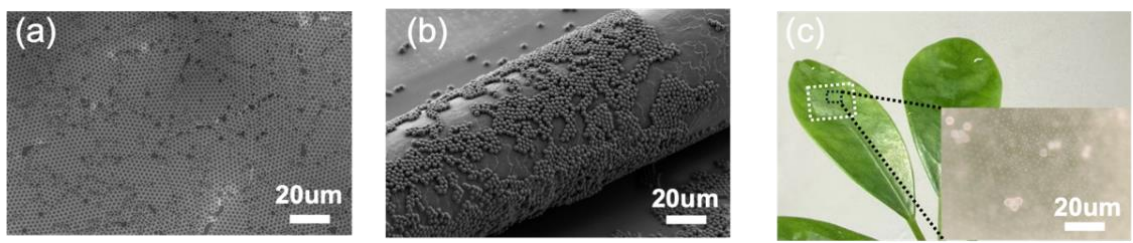

**Figure S14. SEM images showing the transfer performance of 3 µm polystyrene (PS) microspheres via the water-soluble tape transfer printing method.** (a) Single-layer PS microsphere template on the silicon donor substrate. (b) Transferred PS microspheres on hair. (c) Transferred PS microspheres on a leaf surface.